\documentclass{article}
\usepackage[english]{babel}

\usepackage{latexsym}
\usepackage{amssymb}

\newenvironment{tab}{\begin{tabbing}
MMMMM\=aaa\=aaa\=aaa\=aaa\=aaa\=aaa\= \kill}{\end{tabbing}}

\makeatletter
\def\refmystepcounter#1{\stepcounter{#1}\protect\gdef 
\@currentlabel {\csname p@#1\endcsname \csname 
the#1\endcsname}}
\makeatother
\newcounter {tabnr}

\def\Oh{{\cal O}}

\def\omit#1{}
\def\S #1/{\mbox {\textsl{#1}}}
\def\B #1/{\mbox {\textbf{#1}}}
\def\R #1/{\mbox {\textrm{#1}}}
\def\T #1/{\mbox {\texttt{#1}}}

\def\la    {\mbox{$\langle$}}
\def\ra    {\mbox{$\rangle$}}

\def\phi   {{\mbox{$\varphi$}}}
\def\implies{\mbox{$\Rightarrow $}}

\def\Implies{\mbox{$\;\Rightarrow\; $}}
\def\IMPLIES{\mbox{$\quad\Rightarrow\quad $}}
\def\EQ     {\mbox{\quad$\equiv$\quad}}

\def\Land   {\mbox{ $\;\land\;$ }}
\def\LOR    {\mbox{$\quad\lor\quad$}}
\def\Lor    {\mbox{$\;\lor\;$}}

\def\true   {\S true/}
\def\false  {\S false/}
\def\IS     {\mbox{$\quad =\quad $}}

\def\mbreak {\medbreak\noindent}

\def\DOTS   {\mbox{\ .\ .\ }}

\newcounter{theoremcnt}[section]
\newcommand{\qed}{\hfill$\quad\Box$}
\renewcommand{\thetheoremcnt}{\thesection.\arabic{theoremcnt}}

{\begin{trivlist}\refstepcounter{theoremcnt}
\item[]\bf Theorem~\thetheoremcnt.~\rm}{\end{trivlist}}

\begin{document}

\begin {center}
{\Large\bf Lock-free Dynamic Hash Tables \\
with Open Addressing}\\
\mbox{}\\
Gao, H.$^1$, Groote, J.F.$^2$, Hesselink, W.H.$^1$\\
{\small
$^1$ Department of Mathematics and Computing Science, University of 
Groningen,\\
P.O. Box 800, 9700 AV  Groningen, The Netherlands (Email: 
\verb!{hui,wim}@cs.rug.nl!)\\
$^2$ Department of Mathematics and Computing Science, 
Eindhoven University of Technology, P.O. Box 513, 5600 MB  
Eindhoven, The Netherlands and CWI,\\ 
P.O. Box 94079, 1090 GB  Amsterdam, The Netherlands 
(Email: \verb!jfg@win.tue.nl!)}
\end{center}

\begin{abstract}

\noindent 
We present an efficient lock-free algorithm for parallel 
accessible hash tables with open addressing, which promises more 
robust performance and reliability than conventional lock-based 
implementations. ``Lock-free'' means that it is guaranteed 
that always at least one process completes its operation within 
a bounded number of steps. For a single processor architecture 
our solution is as efficient as sequential hash tables. On a 
multiprocessor architecture this is also the case when all 
processors have comparable speeds. The algorithm allows processors 
that have widely different speeds or come to a halt. It can easily be 
implemented using C-like languages and requires on average only 
constant time for insertion, deletion or accessing of elements. 
The algorithm allows the hash tables to grow and shrink when needed. 

Lock-free algorithms are hard to 
design correctly, even when apparently straightforward. Ensuring 
the correctness of the design at the earliest possible stage is a 
major challenge in any responsible system development. In view of 
the complexity of the algorithm, we turned to the interactive 
theorem prover PVS for mechanical support. We employ standard 
deductive verification techniques to prove around 200 invariance 
properties of our algorithm, and describe how this is achieved with
the theorem prover PVS.
\end{abstract}

\mbreak
{\it CR Subject Classification (1991):} 
{\sf D.1 Programming techniques}
{\it Keywords \& Phrases:} {\sf Hash tables, Distributed algorithms, 
Lock-free, Wait-free}

\section{Introduction}
We are interested in efficient, reliable, parallel algorithms. 
The classical synchronization paradigm based on mutual exclusion
is not most suited for this, since mutual exclusion often turns 
out to be a performance bottleneck, and failure of a single process 
can force all other processes to come to a halt. This is the reason 
to investigate lock-free or wait-free concurrent objects, see e.g.\ 
\cite{Bar93,Her91,Her93,HeM92,HG99,KS97,LaMa94,Mic02,RaG02,ShS03,
SuT04,Valois94}. 

\subsubsection*{Lock-free and wait-free objects}

A \emph{concurrent object} is an abstract data type that 
permits concurrent operations that appear to be atomic
\cite{Her91,Lyn96,Valois94}.  The easiest way to implement 
concurrent objects is by means of mutual exclusion, but 
this leads to blocking when the process that holds 
exclusive access to the object is delayed or stops 
functioning.

The object is said to be \emph{lock-free} if any process 
can be delayed at any point without forcing any other 
process to block and when, moreover, it is guaranteed that 
always some process will complete its operation in a 
finite number of steps, regardless of the execution speeds 
of the processes \cite{Bar93,Her93,LaMa94,RaG02,Valois94}.
The object is said to be \emph{wait-free} when it is 
guaranteed that any process can complete any operation in 
a finite number of steps, regardless of the speeds of the 
other processes \cite {Her91}. 

We regard ``non-blocking'' as synonymous to ``lock-free''.  In several
recent papers, e.g. \cite{SuT04}, the term ``non-blocking'' is used
for the first conjunct in the above definition of lock-free. Note that
this weaker concept does not in itself guarantee progress. Indeed,
without real blocking, processes might delay each other arbitrarily
without getting closer to completion of their respective operations.
The older literature \cite{ARJ97,Bar93,HeM92} seems to suggest that
originally ``non-blocking'' was used for the stronger concept, and
lock-free for the weaker one. Be this as it may, we use lock-free 
for the stronger concept.

\subsubsection*{Concurrent hash tables}

The data type of hash tables is very commonly used to efficiently 
store huge but sparsely filled tables. Before 2003, as far as we 
know, no lock-free algorithm for hash tables had been 
proposed. There were general algorithms for arbitrary wait-free 
objects \cite{ABD90,BD89,Bar93,Her91,H95,H96}, but these are not 
very efficient. Furthermore, there are lock-free algorithms for 
different domains, such as linked lists \cite{Valois94}, queues 
\cite{Valois92} and memory management \cite{HeM92,HG99}.

In this paper we present a lock-free algorithm for hash 
tables with open addressing that is in several 
aspects wait-free.  The central idea is that every process 
holds a pointer to a hash table, which is the current one if the 
process is not delayed. When the current hash table is full, a new
hash table is allocated and all active processes join in the 
activity to transfer the contents of the current table to the new 
one. The consensus problem of the choice of a new table is solved 
by means of a test-and-set register. When all processes have left 
the obsolete table, it is deallocated by the last one leaving. This
is done by means of a compare-and-swap register. Measures have been 
taken to guarantee that actions of delayed processes are never 
harmful. For this purpose we use counters that can be incremented 
and decremented atomically. 

After the initial design, it took us several years to establish the 
safety properties of the algorithm. We did this by means of the 
proof assistant PVS \cite{OSR01}. Upon completion of this proof, we 
learned that a lock-free resizable hash table based on chaining was 
proposed in \cite{ShS03}. We come back to this below.

Our algorithm is lock-free and some of the subtasks are wait-free. 
We allow fully parallel \S insertion/, \S assignment/, \S deletion/, 
and \S finding/ of elements. Finding is wait-free, the other three 
are not. The primary cause is that the process executing it may 
repeatedly have to execute or join a migration of the hash table. 
Assignment and deletion are also not wait-free when other processes 
repeatedly assign to the same address successfully.

Migration is called for when the 
current hash table is almost filled. This occurs when the table has 
to grow beyond its current upper bound, but also for maintenance 
after many insertions and deletions. The migration itself is 
wait-free, but, in principle, it is possible that a slow process is 
unable to access and use a current hash table since the current hash 
table is repeatedly replaced by faster processes.

Migration requires subtle provisions, which can be best understood 
by considering the following scenario. Suppose that process $A$ is 
about to (slowly) insert an element in a hash table $H_1$. Before 
this happens, however, a fast process $B$ has performed migration
by making a new hash table $H_2$, and copying the content from 
$H_1$ to $H_2$. If (and only if) process $B$ did not copy the 
insertion of $A$, $A$ must be informed to move to the new hash table,
and carry out the insertion there. Suppose a process $C$ comes into
play also copying the content from $H_1$ to $H_2$. This must be
possible, since otherwise $B$ can stop copying, blocking all 
operations of other processes on the hash table, and thus violating 
the lock-free nature of the algorithm. Now the value inserted by 
$A$ can but need not be copied by both $B$ and/or $C$. This can be 
made more complex by a process $D$ that attempts to replace $H_2$ by 
$H_3$. Still, the value inserted by $A$ should show up exactly once 
in the hash table, and it is clear that processes should carefully 
keep each other informed about their activities on the tables.

\subsubsection*{Performance, comparison, and correctness}

For a single processor architecture our solution is of the same order
of efficiency as sequential hash tables. Actually, only an extra check
is required in the main loop of the main functions, one extra bit
needs to be set when writing data in the hashtables and at some places
a write operation has been replaced by a compare and swap, which is
more expensive.  For ordinary operations on the hashtable, this is the
only overhead and therefore a linear speed up can be expected on
multiprocessor systems. The only place where no linear speed up can be
achieved is when copying the hashtable. Especially, when processes
have widely different speeds, a logaritmic factor may come into play
(see algorithms for the write all problem \cite{Jan00,KS97}). Indeed,
initial experiments indicate that our algorithm is as efficient as
sequential hash tables. It seems to require on average only constant
time for insertion, deletion or accessing of elements.

Some differences between our algorithm and the algorithm of 
\cite{ShS03} are clear. In our algorithm, the hashed values need not 
be stored in dynamic nodes if the address-value pairs (plus one 
additional bit) fit into one word. 
Our hash table can shrink whereas the table of bucket headers in 
\cite{ShS03} cannot shrink. A disadvantage of our algorithm, due 
to its open addressing, is that migration is needed as maintenance 
after many insertions and deletions. 

An apparent weakness of our algorithm is the worst-case space
complexity in the order of $\Oh(P M)$ where $P$ is the number of
processes and $M$ is the size of the table. This only occurs when many
of the processes fail or fall asleep while using the hash
table. Failure while using the hash table can be made less probable by
adequate use of procedure ``releaseAccess''.  This gives a trade-off
between space and time since it introduces the need of a corresponding
call of ``getAccess''.  When all processes make ordinary progress and
the hash table is not too small, the actual memory requirement in
$\Oh(M)$.

The migration activity requires worst-case $\Oh(M^2)$ time for each 
participating process. This only occurs when the migrating processes
tend to choose the same value to migrate and the number of collisions 
is $\Oh(M)$ due to a bad hash function. 
This is costly, but even this is in agreement with 
wait-freedom. The expected amount of work for migration for all 
processes together is $\Oh(M)$ when collisions are sparse,
as should be the case when migrating to a hash table that is 
sufficiently large.

A true problem of lock-free algorithms is that they are hard to 
design correctly, which even holds for apparently straightforward 
algorithms. Whereas human imagination generally suffices to deal 
with all possibilities of sequential processes or synchronized
parallel processes, this appears impossible (at least to us) 
for lock-free algorithms. The only technique that we see fit 
for any but the simplest lock-free algorithms is to prove the 
correctness of the algorithm very precisely, and to verify this 
using a proof checker or theorem prover. 

As a correctness notion, we take that the operations behave the same 
as for `ordinary' hash tables, under some arbitrary linearization
\cite{HeW90} of these operations. So, if a \S find/ is carried out 
strictly after an \S insert/, the inserted element is found. If 
\S insert/ and \S find/ are carried out at the same time, it may be 
that \S find/ takes place before \S insertion/, and it is not 
determined whether an  element will be returned.

Our algorithm contains 81 atomic statements. The structure of our 
algorithm and its correctness properties, as well as the complexity 
of reasoning about them, makes neither automatic nor manual 
verification feasible. We have therefore chosen the higher-order 
interactive theorem prover PVS \cite{FCB00,OSR01} for mechanical 
support. PVS has a convenient specification language and contains 
a proof checker which allows users to construct proofs 
interactively, to automatically execute trivial proofs, and to 
check these proofs mechanically.

\subsubsection*{Overview of the paper}

Section \ref{interface} contains the description of the hash table 
interface offered to the users. The algorithm is presented in 
Section \ref{algorithm}. Section \ref{safety} contains a description 
of the proof of the safety properties of the algorithm: functional 
correctness, atomicity, and absence of memory loss. This proof is 
based on a list of around 200 invariants, presented in Appendix A, 
while the relationships between the invariants are given by a 
dependency graph in Appendix B. 
Progress of the algorithm is proved informally in Section 
\ref{progress}.
Conclusions are drawn in Section \ref{conclusions}.

\section{The interface}\label{interface}

The aim is to construct a hash table that can be 
accessed simultaneously by different processes in such 
a way that no process can passively block another 
process' access to the table. 

A hash table is an implementation of (partial) functions between 
two domains, here called \S Address/ and $\S Value/$. The hash 
table thus implements a modifiable shared variable 
$\T X/: \S Address/\to\S Value/$. 
The domains \S Address/ and \S Value/ both contain special 
default elements $0\in\S Address/$ and $\B null/\in\S Value/$. 
An equality $\T X/(a )=\B null/$ means that no value is currently 
associated with the address $a$. 
In particular, since we never store a value for the address $0$, 
we impose the invariant
\begin {tab}
\> $ \T X/(0)=\B null/$ .
\end {tab}
We use open addressing to keep all elements within the table.
For the implementation of the hash table we require 
that from every value the address it corresponds to is derivable. 
We therefore assume that some function 
$\S ADR/: \S Value/\to\S Address/$ is given with the property that 
\begin{tab} 
\S Ax1:/ \>$~~~v=\B null/~\equiv ~ \S ADR/(v)= 0 $
\end{tab}
Indeed, we need \B null/ as the value corresponding to the undefined 
addresses and use address 0 as the (only) address associated with 
the value \B null/. We thus require the hash table to satisfy the 
invariant
\begin {tab}
\> $ \T X/(a) \ne \B null/\IMPLIES
\S ADR/(\T X/(a))= a$ .
\end {tab}
Note that the existence of \S ADR/ is not a real restriction since 
one can choose to store the pair $(a,v)$ instead of $v$. When $a$ 
can be derived from $v$, it is preferable to store $v$, since that 
saves memory.

There are four principle operations: \S find/, 
\S delete/, \S insert/ and \S assign/. 
The first one is to \S find/ the 
value currently associated with a given address. This 
operation yields $\B null/$ if the address has no associated value. 
The second operation is to \S delete/ the 
value currently associated with a given address. It fails
if the address was empty, i.e.\ $\T X/(a)=\B null/$.
The third operation is to \S insert/ a new value for a given 
address, provided the address was empty. So, note that
at least one out of two consecutive $\S insert/$s for address $a$
must fail, except when there is a $\S delete/$ for address $a$ in
between them. The operation $\S assign/$ does the same
as \S insert/, except that it rewrites the value even if the
associated address is not empty. Moreover, \S assign/ never fails.

We assume that there is a bounded number of processes that 
may need to interact with the hash table. Each process is 
characterized by the sequence of operations
\begin{tab}
\> $(\; \S getAccess/\;;\;
 (\S find/+\S delete/+\S insert/+\S assign/)^*\;;\;
\S releaseAccess/ )^\omega $ 
\end{tab}
A process that needs to access the table, first calls the 
procedure \S getAccess/ to get the current hash table pointer.
It may then invoke the procedures \S find/, \S delete/, \S insert/, 
and \S assign/ repeatedly, in an arbitrary, serial manner.
A process that has access to the table can call \S releaseAccess/ 
to log out. The processes may call these procedures concurrently. 
The only restriction is that every process can do at most one 
invocation at a time. 

The basic correctness conditions for concurrent systems are 
functional correctness and atomicity, say in the sense of 
\cite{Lyn96}, 
Chapter 13. Functional correctness is expressed by prescribing how 
the procedures \S find/, \S insert/, \S delete/, \S assign/ affect 
the value of the abstract mapping \T X/ in relation to the return 
value. Atomicity means that the effect on \T X/ and the return value
takes place atomically at some time between the invocation of the 
routine and its response. 
Each of these procedures has the precondition 
that the calling process has access to the table. In this 
specification, we use auxiliary private variables declared 
locally in the usual way. We give them the suffix $S$ to 
indicate that the routines below are the specifications of the 
procedures. We use angular brackets \la\ and \ra\ to indicate 
atomic execution of the enclosed command.
\begin {tab}
\> $ \B proc /\S find/_S (a:\S Address/\setminus\{0\}) : 
\S Value/ = $\\
\>\>\B local/ $rS:\S Value/$;\\
(fS) \>\>\la~$rS:=\; \T X/(a)\;\ra $;\\
\>\B return/ $rS$.
\end{tab}

\begin {tab}
\> $ \B proc /\S delete/_S (a:\S Address/\setminus\{0\}) : 
\S Bool/ = $\\
\>\>\B local/ $\S sucS/:\S Bool/$;\\
(dS) \>\>\la~$ \S sucS/:=(\T X/(a)\not=\B null/)$ ; \\
\>\>~~$ \B if /\;\S sucS/\; \B\ then /\;
\T X/(a):=\B null/\;\B\ end /\ra$ ;\\
\>\B return/ $\S sucS/$.
\end{tab}

\begin{tab}
\>\B proc/ $\S insert/_S (v:\S Value/\setminus\{\B null/\}) : 
\S Bool/ = $\\
\>\>\B local/ $\S sucS/:\S Bool/\;;\; a:\S Address/:= \S ADR/(v) $ ;\\
(iS) \>\>\la~$ \S sucS/:=(\T X/(a)=\B null/)$ ;\\
\>\>~~$ \B if /\;\S sucS/\; \B\ then /\; \T X/(a):=v\;\B\ end /\ra $ ;\\
\>\B return/ $\S sucS/$.
\end{tab}

\begin {tab}
\>$\B proc /\S assign/_S(v:\S Value/\setminus\{\B null/\})= $\\
\>\>\B local/ $ a:\S Address/:= \S ADR/(v) $ ;\\
(aS) \>\>\la~$\T X/(a):=v \; \ra $ ;\\
\>\B end/.
\end{tab}
Note that, in all cases, we require that the body of the 
procedure is executed atomically at some moment between the 
beginning and the end of the call, but that this moment need not 
coincide with the beginning or end of the call. This is the reason 
that we do not (e.g.) specify \S find/ by the single line 
$\B return /\T X/(a) $.

Due to the parallel nature of our system we cannot use pre and 
postconditions to specify it. For example, it may happen that 
$\S insert/(v)$ returns \true\ while $\T X/(\S ADR/(v))\ne v $ 
since another process deletes $\S ADR/(v)$ between the execution 
of (iS) and the response of \S insert/. 

In Section \ref{primary}, we provide implementations for the 
operations \S find/, \S delete/, \S insert/, \S assign/. 
We prove partial correctness of the implementations by extending them
with the auxiliary variables and commands used in the specification.
So, we regard \T X/ as a shared auxiliary variable and $rS$ 
and $\S sucS/$ as private auxiliary variables; we augment the 
implementations of \S find/, \S delete/, \S insert/, \S assign/ 
with the atomic commands (fS), (dS), (iS), (aS), respectively. 
We prove that each of the four implementations executes 
its specification command always exactly once and that the 
resulting value $r$ or $\S suc/$ of the implementation equals the 
resulting value $rS$ or $\S sucS/$ in the specification. It follows 
that, by removing the 
implementation variables from the combined program, we obtain the 
specification. This removal may eliminate many atomic steps of the 
implementation. This is known as removal of stutterings in TLA 
\cite{Lam94} or abstraction from $\tau$ steps in process algebras.

\section{The algorithm} \label{algorithm}

An implementation consists of $P$ processes along with a set of 
variables, for $P\geq 1$. Each process, numbered from $1$ up to $P$, 
is a sequential program comprised of atomic statements. Actions on 
private variables can be added to an atomic statement, but all 
actions on shared variables must be separated into atomic accesses.
 Since auxiliary variables are only used to facilitate the 
proof of correctness, they can be assumed to be touched instantaneously
without violation of the atomicity restriction. 

\subsection{Hashing}

We implement function \T X/ via hashing with open addressing, cf.\ 
\cite{Knuth3,Wir}. We do not use direct chaining, where colliding 
entries are stored in a secondary list, as is done in \cite {ShS03}.
A disadvantage of open addressing with deletion of elements is that 
the contents of the hash table must regularly be refreshed by copying 
the non-deleted elements to a new hash table. 
As we wanted to be able to resize the hash tables 
anyhow, we consider this less of a burden. 

In principle, hashing is a way to store address-value pairs 
in an array (hash table) with a length much smaller than the 
number of potential addresses. The indices of the array are 
determined by a hash function. In case the hash function maps 
two addresses to the same index in the array there 
must be some method to determine an alternative index. 
The question how to choose a good hash function and how to
find alternative locations in the case of open addressing is 
treated extensively elsewhere, e.g.\ \cite{Knuth3}.

For our purposes it is convenient to combine these two roles in 
one abstract function \S key/ given by:
\begin {tab}
\> $ \S key/(a: \S Address/,\;l:\S Nat/,\;n:\S Nat/):\S Nat/$ ,
\end {tab}
where $l$ is the length of the array (hash table), that satisfies 
\begin{tab}
\S Ax2:/ \>$ 0\leq \S key/(a,l,n) < l$ 
\end{tab}
for all $a$, $l$, and $n$. The number $n$ serves to 
obtain alternative locations in case of collisions: 
when there is a collision, we re-hash until an empty ``slot'' 
(i.e. \B null/) or the same address in the table is found.
The approach with a third argument $n$ is unusual but 
very general. It is more usual to have a function 
\S Key/ dependent on $a$ and $l$, and use a second 
function \S Inc/, which may depend on $a$ and $l$, to use in 
case of collisions. Then our function \S key/ is 
obtained recursively by 
\begin {tab}
\>\+ $ \S key/(a,l,0) = \S Key/(a,l)$ and 
$ \S key/(a,l,n+1) = \S Inc/(a,l,\S key/(a,l,n)) $ .
\end {tab}
We require that, for any address $a$ and any number $l$, 
the first $l$ keys are all different, as expressed in 
\begin {tab}
\S Ax3:/ \> $ 0\leq k < m < l \IMPLIES 
\S key/(a,l,k) \ne \S key/(a,l,m) $ .
\end {tab}

\subsection{Tagging of values}

As is well known \cite{Knuth3}, hashing with open addressing needs 
a special value $\B del/\in \S Value/$ to replace deleted values. 

When the current hash table becomes full, the processes need to reach 
consensus to allocate a new hash table of new size to replace the 
current one. Then all values except \B null/ and \B del/ must be 
migrated to the new hash table. A value that is being migrated 
cannot be simply removed, since the migrating process may stop 
functioning during the migration. Therefore, a value being copied 
must be tagged in such a way that it is still recognizable. This is 
done by the function \S old/. We thus introduce 
an extended domain of values to be called \S EValue/,
which is defined as follows:
\begin{tab}
\>\S EValue/ = $\{ \B del/\}\cup\S Value/\cup
\{\S old/(v)~|~v\in \S Value/\}$
\end{tab}
We furthermore assume the existence of functions
$\S val/: \S EValue/\to\S Value/$ and $\S oldp/: \S EValue/\to\S Bool/$ that satisfy, for all $v\in \S Value/$:
\begin{tab}
\>$\S val/(v)=v$ \>\>\>\>\>\> $\S oldp/(v)=\false$\\
\>$\S val/(\B del/)=\B null/$ \>\>\>\>\>\> $\S oldp/(\B del/)=\false$\\
\>$\S val/(\S old/(v))=v$ \>\>\>\>\>\> $\S oldp/(\S old/(v))=\true$ 
\end{tab}
Note that the \S old/ tag can easily be implemented by 
designating one special bit in the representation of \S Value/.
In the sequel we write $\B done/$ for $\S old/(\B null/)$.
Moreover, we extend the function $\S ADR/$ to domain $\S EValue/$
by $\S ADR/(v)=\S ADR/(\S val/(v))$.

\subsection {Data structure}

A \S Hash table/ is either $\bot$, indicating the absence of a 
hash table, or it has the following structure:
\begin{tab}
\>$\T size, bound, occ, dels/ :\S Nat/$;\\
\>$\T table/:\B array / 0\DOTS\T size-1 /\B of/ \S\ EValue/$.
\end{tab}
The field \T size/ indicates the size of the hash table, 
\T bound/ the maximal number of places that
can be occupied before refreshing the table. Both
are set when creating the table and remain constant. The 
variable \T occ/ gives the number of occupied positions in
the table, while the variable \T dels/ gives the number of deleted 
positions. If $h$ is a pointer to a hash table, we write 
$h.\T size/$, $h.\T occ/$, $h.\T dels/$ and 
$h.\T bound/$ to access these fields of the hash table. We write 
$h.\T table/[i]$ to access the $i^{\rm th}$ \S EValue/ in 
the table.

Apart from the \S current/ hash table, which is the main representative
of the variable \T X/, we have to deal with \S old/ hash tables, which 
were in use before the current one, and \S new/ hash tables, which can be
 created after the current one.
 
We now introduce data structures that are used by the processes
to find and operate on the hash table and allow to
delete hash tables that are not used anymore. The basic idea is
to count the number of processes that are using a hash table, 
by means of a counter \T busy/. The hash table can be thrown away when
\T busy/ is set to $0$. An important observation is that \T busy/ 
cannot be stored as part of the hash table, in the same way as 
the variables \T size/, \T occ/ and \T bound/ above. The reason 
for this is that a process can attempt to access the current 
hash table by increasing its \T busy/ counter. However, just 
before it wants to write the new value for \T busy/ it falls 
asleep. When the process wakes up the hash table might have been 
deleted and the process would be writing at a random place in 
memory.

This forces us to use separate arrays \T H/ and \T busy/ to store 
the pointers to hash tables and the \T busy/ counters. There
can be $2P$ hash tables around, because each process can
simultaneously be accessing one hash table and attempting to
create a second one. The arrays below are shared variables.

\begin {tab}
\>$\T H/:\B array /1\DOTS 2P \B\ of pointer to /\S Hashtable/$ ;\\
\>$\T busy/ :\B array /1\DOTS 2P \B\ of / \S Nat/$ ;\\
\>$\T prot/:\B array /1\DOTS 2P \B\ of / \S Nat/$ ;\\
\>$\T next/:\B\ array /1\DOTS 2P \B\ of /0\DOTS 2P$ .
\end {tab}
As indicated, we also need arrays \T prot/ and \T next/. The variable 
$\T next/[i]$ points to the next hash table to which the 
contents of hash table $\T H/[i]$ is being copied. 
If $\T next/[i]$ equals $0$, this means that there is no next 
hash table. The variable $\T prot/[i]$
is used to guard the variables $\T busy/[i]$, $\T next/[i]$ and
$\T H/[i]$ against being reused for a new table, before
all processes have discarded them.

We use a shared variable \T currInd/ to hold the index
of the currently valid hash table:
\begin {tab}
\>$\T currInd/:1\DOTS 2P$ .
\end {tab}
Note however that after a process copies $\T currInd/$ 
to its local memory, other processes may create a new 
hash table and change $\T currInd/$ to point to that one.

It is assumed that initially $\T H/[1]$ is pointing to 
some hash table. The other initial values of the shared 
variables are given by
\begin{tab}
\> $ \T currInd/ = \T busy/[1] = \T prot/[1] = 1 $ ,\\
\> $\T H/[i] = \T busy/[i] = 
\T prot/[i] = 0 $ \ for all $i\ne 1$ ,\\
\> $ \T next/[i] = 0 $ \ for all $i$.
\end{tab}

\subsection{Primary procedures} \label{primary}

We first provide the code for the primary procedures, which match
directly with the procedures in the interface. Every process has 
a private variable 
\begin{tab}
\>$\S index/:1\DOTS 2P$;
\end{tab}
containing what it regards as the currently active hash table.
At entry of each primary procedure, 
it must be the case that the variable $\T H/[\S index/]$
contains valid information. 
In section \ref{Mem}, we provide procedure \S getAccess/ with 
the main purpose to guarantee this property. 
When \S getAccess/ has been called, the system is obliged to 
keep the hash table at \S index/ stored in memory, even if there 
are no accesses to the hash table using any of the primary 
procedures. A procedure \S releaseAccess/ is provided to 
release resources, and it should be called whenever the 
process will not access the hash table for some time.

\subsubsection{Syntax}

We use a syntax analogous to Modula-3 \cite{Har92}. We use $:=$ for 
the assignment. We use the C--operations \T ++/ and 
\T --/ for atomic increments and decrements. 
The semicolon is a separator, not a terminator. 
The basic control mechanisms are
\begin{tab}
\> \B loop .. end / is an infinite loop, terminated by 
\B exit/ or \B return/\\
\> \B while .. do .. end / and \B repeat .. until .. / are 
ordinary loops\\
\> \B if .. then .. \{elsif ..\} [else ..] end / is the 
conditional\\
\> \B case .. end / is a case statement.
\end{tab}
Types are slanted and start with a capital. Shared variables 
and shared data elements are in typewriter font. Private variables 
are slanted or in math italic. 

\subsubsection{The main loop}
We model the clients of the hash table in the following loop. This 
is not an essential part of the algorithm, but it is needed in the
PVS description, and therefore provided here.
\begin {tab}
\>\B loop/\\
0: \>\> $ \S getAccess/()$ ;\\
\>\>\B loop/\\
1: \>\>\>$\S choose call; /\B case /call\ \B of /$\\
\>\>\>\> $(f,a)\ \B with /\ a\neq 0\rightarrow \S find/(a)$\\
\>\>\>\> $(d,a)\ \B with /\ a\neq 0\rightarrow \S delete/(a)$\\
\>\>\>\> $(i,v)\ \B with /\ v\neq \B null/\rightarrow 
\S insert/(v)$\\
\>\>\>\> $(a,v)\ \B with /\ v\neq \B null/\rightarrow 
\S assign/(v)$\\
\>\>\>\> $(r)\ \rightarrow \S releaseAccess/(\S index/);\ \B exit/$\\
\>\>\>\B end/\\
\>\>\B end/\\
\>\B end/
\end{tab}
The main loop shows that each process repeatedly invokes its four 
principle operations with correct arguments in an arbitrary, serial 
manner. Procedure \S getAccess/ has to provide the client with a 
protected value for \S index/. Procedure \S releaseAccess/ 
releases this value and its protection. Note that \B exit/ means 
a jump out of the inner loop.

\subsubsection{Procedure \S find/}

Finding an address in a hash table with open addressing requires 
a linear search over the possible hash keys until the address or 
an empty slot is found. The kernel of procedure \S find/ is 
therefore:
\begin {tab}
\>\+ $ n:= 0 $ ;\\
$ \B repeat /\; r:=h.\T table/[\S key/(a, l, n)] \;;\quad
n\T ++/ $ ;\\
$ \B until / \; r= \B null/\Lor a = \S ADR/(r) $ ;
\end{tab}
The main complication is that, when the process encounters an entry 
\B done/ (i.e. \S old/(\B null/)), it has to join the migration 
activity by calling \S refresh/. 

Apart from a number of special commands, we group statements
such that at most one shared variable is accessed and label these
`atomic' statements with a number. The labels are chosen identical 
to the labels in the PVS code, and therefore not completely 
consecutive.

In every execution step, one of the processes proceeds from 
one label to a next one. The steps are thus treated as 
atomic. The atomicity of steps that refer to shared variables 
more than once is emphasized by enclosing them in angular 
brackets. Since procedure calls only modify private control data, 
procedure headers are not always numbered themselves, but their bodies 
usually have numbered atomic statements.
\begin {tab}
\>$ \B proc /\S find/ (a:\S Address/\setminus\{0\}) : 
\S Value/= $\\
\>\>\B local/ $ r:\S EValue/$ ; $ n,l:\S Nat/\;;\; 
h:\B pointer to /\S Hashtable/$ ;\\
5: \>\> $h:=\T H/[\S index/]\;;\; n:=0 \;;\;\{ \S cnt/:=0\}$ ;\\
6: \>\> $l:=h.\T size/$ ;\\
\>\>\B repeat/\\
7: \>\>\>\la~$r:=h.\T table/[\S key/(a, l, n)] $ ;\\
\>\>\>~~$\{ \B\ if /r= \B null/\Lor a=\S ADR/(r)\B\ then /\;
\S cnt/\T ++/\;;\;\R (fS)/ \;\B\ end /\}\;\ra $ ;\\
8: \>\>\> \B if / $r=\B done/\;\B\ then /$\\
\>\>\>\> $ \S refresh/()$ ;\\
10: \>\>\>\>$ h:=\T H/[\S index/]\;;\; n:=0$ ;\\
11: \>\>\>\>$l:=h.\T size/$ ;\\
\>\>\>$ \B else /\; n\T ++/\; \B\ end/$ ;\\
13: \>\>\B until/ $r= \B null/\Lor a=\S ADR/(r) $ ;\\
14: \>\B return/ $\S val/(r)$ .
\end{tab}

In order to prove correctness, we add between braces 
 instructions that only modify auxiliary variables, like the 
specification variables \T X/ and $rS$ and other auxiliary 
variables to be introduced later.  
The part between braces is comment for the implementation, it 
only serves in the proof of correctness. The private auxiliary 
variable \S cnt/ of type \S Nat/ counts the number of times 
(fS) is executed and serves to prove that (fS) is executed 
precisely once in every call of \S find/. 

This procedure matches the code of an ordinary find in a hash table 
with open addressing, except for the code at the 
condition $r=\B done/$. This code is needed for the case that 
the value at address $a$ has been copied, in which case the new 
table must be located. Locating the new table is carried out 
by the procedure \S refresh/, which is discussed in Section 
\ref{Mem}. In line 7, the accessed hash table should be 
valid (see invariants \S fi4/ and \S He4/ in Appendix A).
After \S refresh/ the local variables $n$, $h$ and 
$l$ must be reset, to restart the search in the new hash table.
If the procedure terminates, the specifying atomic command (fS) 
has been executed precisely once (see invariant \S Cn1/) and the 
return values of the specification and the implementation are equal 
(see invariant \S Co1/). If the operation succeeds, 
the return value must be a valid entry currently associated with 
the given address in the current hash table. It is not evident but 
it has been proved that the linear search of the process executing 
\S find/ cannot be violated by other processes, i.e. no other 
process can \S delete/, \S insert/, or \S rewrite/ an entry 
associated with the same address (as what the process is looking 
for) in the region where the process has already searched. 

We require that every valid hash table contains at least one entry 
\B null/ or \B done/. Therefore, the local variable $n$ in the 
procedure \S find/ never goes beyond the size of the hash table 
(see invariants \S Cu1/, \S fi4/, \S fi5/ and axiom \S Ax2/).
When the \T bound/ of the new hash table is tuned properly before 
use (see invariants \S Ne7/, \S Ne8/), 
the hash table will not be updated too frequently, 
and termination of the procedure \S find/ can be guaranteed. 

\subsubsection{Procedure \S delete/}

To some extent, deletion is similar to finding. Since $r$ is a 
local variable to the procedure \S delete/, we regard 18a and 18b 
as two parts of atomic instruction 18. 
 If the entry is found in the table, then at line 18b this entry 
is overwritten with the designated element \B del/.

\begin {tab}
\> $ \B proc /\S delete/ (a:\S Address/\setminus\{0\}) : 
\S Bool/ = $\\
\>\>\B local/ $ r:\S EValue/\;;\; k,l,n:\S Nat/ $ ;\\
\>\>\> $ h:\B pointer to /\S Hashtable/\;;\;\S suc/:\S Bool/$ ;\\
15: \>\> $h:=\T H/[\S index/]\;;\;\S suc/:=\; \false\;;\; 
\{\S cnt/:=0\} $ ;\\
16: \>\>$l:=h.\T size/\;;\; n:=0$ ;\\
\>\>\B repeat/\\
17: \>\>\> $k:=\S key/(a, l, n) $ ;\\
\>\>\> \la~$ r:=h.\T table/[k] $ ;\\
\>\>\>~~$\{\;\B  if /r=\B null/\;\B\ then /\;
\S cnt/\T ++/\;;\; \R (dS) /\B\ end /\}\;\ra $ ;\\
18a: \>\>\> $ \B if /\;\S oldp/(r)\;\B\ then / $\\
\>\>\>\>  $ \S refresh/()$ ;\\
20: \>\>\>\> $ h:=\T H/[\S index/]$ ; \\
21: \>\>\>\> $ l:=h.\T size/\;;\; n:=0$ ;\\
\>\>\> $ \B elsif / \; a=\S ADR/(r)\;\B\ then / $\\
18b:\>\>\>\> \la~$ \B if /\; r=h.\T table/[k] \;\B\ then /$\\
\>\>\>\>\>\>~~$\S suc/:= \true \;;\; h.\T table/[k] :=\B del/ $ ; \\
\>\>\>\>\>\>~~$\{\; \S cnt/\T ++/\;;\;\R (dS) /\;;\;
\T Y/[k] :=\B del/\ \}$\\
\>\>\>\>~~\B end /\ra\ ;\\
\>\>\> $\B else /\; n\T++/ \;\B\ end/ $ ; \\
\>\> $ \B until /\; \S suc/ \Lor r = \B null/ $ ;\\
25: \>\> $ \B if /\; \S suc/\; \B\ then /\; 
h.\T dels/ \T ++/\;\B\ end/$ ;\\
26: \> \B return/ \S suc/ .
\end{tab}

The repetition in this procedure has two ways to terminate.
Either deletion fails with $r=\B null/$ in 17, or deletion succeeds 
with $r=h.\T table/[k]$ in 18b. In the latter case, we have in one 
atomic statement a double access of the shared variable 
$h.\T table/[k]$. This is a so-called compare\&swap instruction. 
Atomicity is needed here to preclude interference. The specifying 
command (dS) is executed either in 17 or in 18b, and it is executed 
precisely once (see invariant \S Cn2/), since in 18b the guard 
$a=\S ADR/(r)$ implies $r\ne\B null/$ (see invariant 
\S de1/ and axiom \S Ax1/). 

In order to remember the address from the value rewritten 
to \B done/ after the value is 
being copied in the procedure \S moveContents/, in 18, we 
introduce a new auxiliary shared variable \T Y/ of type array 
of \S EValue/, whose contents 
equals the corresponding contents of the current hash table 
almost everywhere except that the values it contains are not 
tagged as \S old/ or rewritten as \B done/ (see invariants 
\S Cu9/, \S Cu10/).
 
Since we postpone the increment of \S h/.\T dels/ until line 25, 
the field \T dels/ is a lower bound of the number of positions 
deleted in the hash table (see invariant \S Cu4/). 

\subsubsection{Procedure \S insert/}

The procedure for insertion in the table is given below. 
Basically, it is the standard algorithm for insertion in a 
hash table with open addressing. Notable is line 28 where 
the current process finds that the current hash table too full, 
and orders a new table to be made.  
We assume that $h.\T bound/$ is a number less than 
$h.\T size/$ (see invariant \S Cu3/), which is tuned for optimal 
performance. 

Furthermore, in line 35, it can be detected that values in the 
hash table have been marked \S old/, which is a sign that 
hash table $h$ is outdated, and the new hash table must 
be located to perform the insertion.
\begin {tab}
\>$ \B proc /\S insert/ (v:\S Value/\setminus\{\B null/\}) : 
\S Bool/ = $\\
\>\> $ \B local / r:\S EValue/\;;\; k,l,n:\S Nat/\;;\;
h:\B pointer to /\S Hashtable/$ ;\\
\>\>\> $\S suc/:\S Bool/\;;\; a:\S Address/:= \S ADR/(v) $ ;\\
27: \>\>$h:=\T H/[\S index/] \;;\; \{\S cnt/:=0\} $ ;\\
28: \>\>\B if/ $h.\T occ/ > h.\T bound/\B\ then /$\\
\>\>\> $ \S newTable/()$ ;\\
30: \>\>\> $h:=\T H/[\S index/]\; \B\ end/$ ;\\
31: \>\> $ n:=0\;;\; l:=h.\T size/\;;\; \S suc/:= \false$ ;\\
\>\> \B repeat /\\
32: \>\>\>$ k:=\; \S key/(a,l,n) $ ; \\
33: \>\>\>\la~$ r:=h.\T table/[k] $ ; \\
\>\>\>~~$ \{\;\B if /\;a=\S ADR/(r)\;
\B\ then /\; \S cnt/\T ++/\;;\;\R\ (iS) /\B\ end /\}\;\ra $ ;\\
35a: \>\>\>$ \B if /\;\S oldp/(r)\; \B\ then/ $ \\
\>\>\>\> $ \S refresh/()$ ;\\
36: \>\>\>\> $h:=\T H/[\S index/]$ ;\\
37: \>\>\>\> $n:=0\;;\; l:=h.\T size/$ ;\\
\>\>\> $ \B elsif / \; r = \B null/ \B\ then /$\\
35b: \>\>\>\> \la~$ \B if /\; h.\T table/[k] = \B null/
\;\B\ then/$\\
\>\>\>\>\> $  \S suc/:=\true \;;\; h.\T table/[k] :=v $ ; \\
\>\>\>\>\> $ \{\; \S cnt/\T ++/\;;\;\R (iS)/\;;\;
\T Y/[k] :=v\ \} $ \\
\>\>\>\>~~\B end /\ra\ ;\\
\>\>\> $\B else /\; n\T ++/\; \B\ end / $ ;\\
\>\> $ \B until /\;\S suc/\Lor a=\S ADR/(r)$ ;\\
41: \>\> $ \B if /\; \S suc/\; \B\ then /\; 
h.\T occ/ \T ++/\;\B\ end/$ ;\\
42: \> $ \B return /\; \S suc/$ .
\end{tab}

Instruction 35b is a version of compare\&swap. Procedure \S insert/
terminates successfully when the insertion to an empty slot is
completed, or it fails when there already exists an entry with the
given address currently in the hash table (see invariant \S Co3/ and
the specification of \S insert/).

\subsubsection{Procedure \S assign/}

Procedure \S assign/ is almost the same as \S insert/ except that 
it rewrites an entry with a given value even when the associated 
address is not empty. We provide it without further comments.
\begin {tab}
\>$ \B proc /\S assign/ (v:\S Value/\setminus\{\B null/\})
\ =$\\
\>\>\B local/ $ r:\S EValue/\;;\; k,l,n:\S Nat/\;;\;
h:\B pointer to /\S Hashtable/ $ ;\\
\>\>\> $ \S suc/:\S Bool/\;;\; a:\S Address/:=\S ADR/(v) $ ;\\
43: \>\>$h:=\T H/[\S index/]\;;\; \S cnt/:=0$;\\
44: \>\> $\B if /\; h.\T occ/> h.\T bound/\;\B\ then /$\\
\>\>\> $ \S newTable/()$ ;\\
46: \>\>\> $ h:=\T H/[\S index/]\;\B\ end/$ ;\\
47: \>\> $ n:=0 \;;\; l:=h.\T size/\;;\; \S suc/:= \false $ ;\\
\>\>\B repeat/\\
48: \>\>\> $ k:=\; \S key/(a,l,n)$ ;\\
49: \>\>\> $ r:=h.\T table/[k]$ ;\\
50a: \>\>\> $ \B if /\; \S oldp/(r) \;\B\ then/ $\\
\>\>\>\> $ \S refresh/()$ ;\\
51: \>\>\>\> $ h:=\T H/[\S index/]$ ;\\
52: \>\>\>\> $n:=0\;;\; l:=h.\T size/$ ;\\
\>\>\> $ \B elsif /\; r=\B null/\Lor a=\S ADR/(r) \;
\B\ then /$\\
50b: \>\>\>\> \la~$\B if /\;h.\T table/[k] = r\;\B\ then/$\\
\>\>\>\>\>$ \S suc/:=\true \;;\;
h.\T table/[k] := v\;;\; $\\
\>\>\>\>\> $\{\; \S cnt/\T ++/\;;\; \R\ (aS) /;\ \T Y/[k]:=v\ \}$ \\
\>\>\>\>~~\B end /\ra \\
\>\>\>$ \B else /\; n\T ++/\;\B\ end/$ ;\\
\>\> $ \B until /\S suc/ $ ;\\
57: \>\> $ \B if /\; r=\B null/\;\B\ then /\;
h.\T occ/\T ++/\;\B\ end/ $ ; \\
\>\B end/.
\end{tab}

\subsection {Memory management and concurrent migration}
\label{Mem}

In this section, we provide the public procedures \S getAccess/ 
and \S releaseAccess/ and the auxiliary procedures \S refresh/ 
and \S newTable/ which are responsible for allocation and 
deallocation. We begin with the treatment of memory by providing 
a model of the heap.

\subsubsection {The model of the heap}
We \emph{model} the \T Heap/ as an infinite array of 
hash tables, declared and initialized in the following way:
\begin{tab}
\>\+ $\T Heap/:\B\ array /\S Nat/\B\ of /\S Hashtable/:=([\S Nat/]\bot)$ ;\\
$ \T H/_-\T index/:\S Nat/:= 1 $ .
\end{tab}
So, initially, $\T Heap/[i] = \bot$ for all indices $i$. 
The indices of array \T Heap/ are the pointers to hash tables. We 
thus simply regard \B pointer to/ \S Hashtable/ as a synonym of 
\S Nat/. 
Therefore, the notation $h.\T table/$ used elsewhere in the paper 
stands for $\T Heap/[h].\T table/$. Since we reserve 0 (to be 
distinguished from the absent hash table $\bot$ and the absent 
value \B null/) for the null pointer (i.e. $\T Heap/[0]=\bot$, 
see invariant \S He1/), we initialize \T H/$_-$\T index/, which is 
the index of the next hash table, to be 1 instead of 0. Allocation 
of memory is modeled in
\begin{tab}
\>\+\+ $ \B proc /\S allocate/(s, b:\S Nat/): \S Nat/ =$ \\
\la~$\T Heap/ [\T H/_-\T index/]:=\R\ blank hash table with /
\T size/= s, \T bound/=b,$\\
\>\>\>\> $ \T occ/=\T dels/=0$ ;\\
~~$\T H/_-\T index/\T ++/\ \ra $ ;\-\\
$\B return /\T H/_-\T index/ $ ;
\end{tab}
We assume that $\S allocate/$ sets all values in the hash table 
$\T Heap/[\T H/_-\T index/]$ to $\B null/$, and also sets its fields 
$\T size/$ and $\T bound/$ as specified. 
The variables \T occ/ and \T dels/ are set to 0 because 
the hash table is completely filled with the value \B null/.
 
Deallocation of hash tables is modeled by
\begin{tab}
\>\+ $ \B proc /\S deAlloc/(h:\S Nat/) =$ \\
\>\la~$ \B assert /\T Heap/[h]\ne\bot\;;\quad
\T Heap/[h]:=\bot\ \ra $ \\
\B end/ .
\end{tab}
The \B assert/ here indicates the obligation to prove 
that \S deAlloc/ is called only for allocated memory.

\subsubsection{Procedure \S getAccess/}
The procedure \S getAccess/ is defined as follows.
\begin {tab}
\>$\B proc /\S getAccess/() = $ \\
\>\>$ \B loop/$\\
59: \>\>\> $ \S index/ :=\; \T currInd/$;\\
60: \>\>\> $ \T prot/[\S index/]\T ++/$ ;\\
61: \>\>\> $ \B if / \; \S index/ =\T currInd/\; \B\ then /$\\
62: \>\>\>\> $ \T busy/[\S index/]\T++/ $ ;\\
63: \>\>\>\> $ \B if /\; \S index/ =\T currInd/\; 
\B\ then return/ $ ;\\
\>\>\>\> $ \B else /\; \S releaseAccess/(\S index/)\;\B\ end/$ ;\\
65: \>\>\> $ \B else /\;\T prot/[\S index/]\T --/\;\B\ end/$ ;\\
\>\>\B end/\\
\>\B end/.
\end {tab}
This procedure is a bit tricky. When the process reaches line 62,
the \S index/ has been protected not to be used for creating a 
new hash table in the procedure \S newTable/ (see invariants \S pr2/, 
\S pr3/ and \S nT12/).

The hash table pointer $\T H/[\S index/]$ must contain the valid 
contents after the procedure \S getAccess/ returns (see invariants 
\S Ot3/, \S He4/). So, in line 62, \T busy/ is increased, 
guaranteeing that the hash table will not inadvertently be destroyed 
(see invariant \S bu1/ and line 69).
Line 63 needs to check the \S index/ again in case that instruction 
62 has the precondition that the hash table is not valid. Once some 
process gets hold of one hash table after calling \S getAccess/, no 
process can throw it away until the process releases it (see 
invariant \S rA7/). 
 
\subsubsection{Procedure \S releaseAccess/}\label{deallocate}
The procedure \S releaseAccess/ is given by
\begin {tab}
\>$\B proc /\S releaseAccess/(i:1\DOTS 2P) = $\\
\>\>\B local/ $h:\B pointer to /\S Hashtable/$ ;\\
67: \>\>$ \S h/ :=\; \T H/[i] $ ;\\
68: \>\>$ \T busy/[i]\T --/$ ;\\
69: \>\>$\B if /\;\S h/\ne 0\Land \T busy/[i] = 0\; \B\ then/$\\
70: \>\>\>$\la~\B if /\T H/[i] = h\; \B\ then /\;\T H/[i]:= 0$ ; \ra\\
71: \>\>\>\>$ \S deAlloc/(h)$ ;\\
\>\>\>~~\B end/ ;\\
\>\> \B end/ ;\\
72: \>\> $ \T prot/[i]\T --/$ ;\\
\>\B end/.
\end {tab}
The test $h\ne 0$ at 69 is necessary since it is possible that $h=0$ at the 
lines 68 and 69. This occurs e.g. in the following scenario. 
Assume that process $p$ is at line 62 with $\S index/ \ne\T currInd/$, while
the number $i=\S index/$ satisfies $\T H/[i] = 0$ and $\T busy/[i]=0$.
Then process $p$ increments $\T busy/[i]$, calls $\S releaseAccess/(i)$, 
and arrives at 68 with $h=0$.

Since \S deAlloc/ in line 71 accesses a shared variable, we have
separated its call from 70.  The counter $\T busy/[i]$ is used to
protect the hash table from premature deallocation. Only if \T
busy/[i]=0, \T H/[i] can be released.  The main problem of the design
at this point is that it can happen that several processes
concurrently execute \S releaseAccess/ for the same value of $i$, with
interleaving just after the decrement of $\T busy/[i]$. Then they all
may find $\T busy/[i]=0$. Therefore, a bigger atomic command is needed
to ensure that precisely one of them sets $\T H/[i]$ to $0$ (line 70)
and calls \S deAlloc/.  Indeed, in line 71, \S deAlloc/ is called only
for allocated memory (see invariant \S rA3/). The counter $\T
prot/[i]$ can be decreased since position $i$ is no longer used by
this process.

\subsubsection{Procedure \S newTable/}

When the current hash table has been used for some time, some 
actions of the processes may require replacement of this hash 
table. Procedure \S newTable/ is called when the number of 
occupied positions in 
the current hash table exceeds the \T bound/ (see lines 28, 44). 
Procedure \S newTable/ tries to allocate a new hash table as the 
successor of the current one.
If several processes call \S newTable/ concurrently, they need to 
reach consensus on the choice of an index for the next hash table 
(in line 84). A newly allocated hash table that will not be used 
must be deallocated again.
\begin {tab}
\>$\B proc /\S newTable/() = $ \\
\>\>\B local/ $ i:1\DOTS 2P\;;\;b , bb:\S Bool/$ ;\\
77: \>\> $ \B while/\; \T next/[\S index/] = 0\; \B\ do/ $\\
78: \>\>\> $ \B choose /i \in 1\DOTS 2P$ ;\\
\>\>\> \la~$b:=(\T prot/[i] = 0) $ ;\\
\>\>\>~~$ \B if /\;b\;\B\ then / 
\T prot/[i]:=1\;\B\ end / \ra $ ;\\
\>\>\> $\B if /\;b\;\B\ then / $ \\
81: \>\>\>\> $ \T busy/[i]:=1$ ;\\
82: \>\>\>\> $\B choose /
\S bound/ > \T H/[\S index/].\T bound/-\T H/[\S index/].\T dels/+2P $ ;\\
\>\>\>\> $\B choose /\S size/> \S bound/+2P $ ;\\
\>\>\>\>$\T H/[i] :=\S allocate/(\S size/, \S bound/)\;; $\\
83: \>\>\>\> $ \T next/[i]:=0 $ ;\\
84: \>\>\>\> \la~$ bb:=(\T next/[\S index/] = 0)$ ;\\
\>\>\>\>~~$ \B if /\; bb\; \B\ then /\; 
\T next/[\S index/] :=i\ \B\ end /\ra $ ;\\
\>\>\>\>$ \B if /\; \neg bb\; \B\ then /\;
\S releaseAccess/(i)\; \B\ end /$ ;\\
\>\> \B end end/ ;\\
\>\>$\S refresh/()$ ;\\
\>\B end/ .
\end {tab}
In command 82, we allocate a new blank hash table (see invariant 
\S nT8/), of which the \T bound/ is set greater than 
$\T H/[\S index/].\T bound/-\T H/[\S index/].\T dels/+2P$ 
in order to avoid creating a too small hash table (see invariants 
\S nT6/, \S nT7/). 
 
We require the \T size/ of a hash table to be more than 
$\T bound/+2P$ because of the following scenario: $P$ processes find 
``$h.\T occ/>h.\T bound/$'' at line 28 and call \S newtable/, 
\S refresh/, \S migrate/, \S moveContents/ and \S moveElement/ one 
after the other. 
After moving some elements, all processes but process $p$ sleep at 
line 126 with $b_{\it mE}=\true$ ($b_{\it mE}$ is the local variable 
$b$ of procedure \S moveElement/). Process $p$ continues the 
migration and updates the new current index when the migration 
completes. Then, process $p$ does several insertions to let the 
\T occ/ of the current hash table reach one more than its \T bound/. 
Just at that moment, $P-1$ processes wake up, increase the \T occ/ 
of the current hash table to be $P-1$ more, and return to line 30. 
Since $P-1$ processes insert different values in the hash table, 
after $P-1$ processes finish their insertions, the \T occ/ of the 
current hash table reaches $2P-1$ more than its \T bound/. 

It may be useful to make \T size/ larger than $\T bound/ + 2P $ to 
avoid too many collisions, e.g. with a constraint 
$\T size/\geq\alpha\cdot\T bound/$ for some $\alpha >1$.
If we did not introduce \T dels/, every migration would force the 
sizes to grow, so that our hash table would require unbounded space 
for unbounded life time. We introduced \T dels/ to avoid this.

Strictly speaking, instruction 82 inspects one shared variable, 
$\T H/[\S index/]$, and modifies three other shared variables, viz. 
$\T H/[i]$, $\T Heap/[\T H/_-\T index/]$, and $\T H/_-\T index/$. 
In general, we split such multiple shared variable accesses in 
separate atomic commands. Here the accumulation is harmless, since 
the only possible interferences are with other allocations at line 
82 and deallocations at line 71. In view of the invariant \S Ha2/, 
all deallocations are at pointers $h<\T H/_-\T index/$. 
Allocations do not interfere because they contain the 
increment $\T H/_-\T index/$++ (see procedure \S allocate/).

The procedure \S newTable/ first searches for a free index $i$, say 
by round robin. We use a nondeterministic choice.
Once a free index has been found, a hash table is allocated and 
the index gets an indirection to the allocated address. Then the 
current index gets a \T next/ pointer to the new 
index, unless this pointer has been set already.

The variables $\T prot/[i]$ are used primarily as 
counters with atomic increments and 
decrements. In 78, however, we use an atomic test-and-set 
instruction.
Indeed, separation of this instruction in two atomic
instructions is incorrect, since that would allow
two processes to grab the same index $i$ concurrently.

\subsubsection{Procedure \S migrate/}
After the choice of the new hash table, the procedure \S migrate/ 
serves to transfer the contents in the current hash table to the new
hash table by calling a procedure \S moveContents/ and to update the 
current hash table pointer afterwards. Migration is complete when at 
least one of the (parallel) calls to \S migrate/ has terminated.

\begin {tab}
\>$ \B proc /\S migrate/() = $ \\
\>\>\B local/ $i:0\DOTS 2P$; $h:\B pointer to /\S Hashtable/\;;\;
b:\S Bool/$ ;\\
94: \>\>$ i:=\T next/[\S index/]$;\\
95: \>\> $ \T prot/[i]\T ++/$ ;\\
97: \>\>$ \B if /\;\S index/\ne\T currInd/\;\B\ then /$\\
98: \>\>\>$ \T prot/[i]\T --/$ ;\\
\>\> \B else / \\
99: \>\>\>$ \T busy/[i] \T ++/$ ;\\
100: \>\>\>$h:=\T H/[i]$ ;\\
101: \>\>\>$ \B if /\; \S index/=\T currInd/\;\B\ then / $\\
\>\>\>\>$ \S moveContents/(\T H/[\S index/], h) $ ;\\
103: \>\>\>\>\la~$b:=(\T currInd/=\S index/)$ ;\\
\>\>\>\>~~$\B if /\; b \;\B\ then /\;\T currInd/:=i$ ;
\ $\{\T Y/:=\T H/[\S i/].\T table/\ \}$\\
\>\>\>\>~~$\B end /\ra$ ;\\
 \>\>\>\>$ \B if /\; b\; \B\ then /$ \\
104: \>\>\>\>\>$ \T busy/[\S index/] \T --/$ ;\\
105: \>\>\>\>\>$ \T prot/[\S index/] \T --/$ ;\\
\>\>\> \B end end/ ;\\
\>\>\> $\S releaseAccess/(i)$ ;\\
\>\B end end /.
\end {tab}

According to invariants \S mi4/ and \S mi5/, it is an invariant that 
$i=\T next/(\S index/)\neq 0$ holds after instruction 94.

Line 103 contains a compare\&swap instruction to update the current 
hash table pointer when some process finds that the migration is 
finished while \T currInd/ is still identical to its \S index/, which 
means that $i$ is still used for the next current hash table (see 
invariant \S mi5/). The increments of $\T prot/[i]$ and $\T busy/[i]$ 
here are needed to protect the next hash table. The decrements serve
to avoid memory loss.

\subsubsection{Procedure \S refresh/}
In order to avoid that a delayed process starts migration of an old 
hash table, we encapsulate \S migrate/ in \S refresh/ in the 
following way.
\begin {tab}
\>$ \B proc /\S refresh/() = $\\
90: \>\>$ \B if /\; \S index/\ne\T currInd/\; \B\ then/ $\\
\>\>\>$ \S releaseAccess/(\S index/) $ ;\\
\>\>\>$ \S getAccess/() $ ;\\ 
\>\>$ \B else /\;\S migrate/()\;\B\ end/$ ;\\
\>\B end/.
\end {tab}
When \S index/ is outdated, the process needs to call
\S releaseAccess/ to abandon its hash table and \S getAccess/ to acquire 
the present pointer to the current hash table.
Otherwise, the process can join the migration.

\subsubsection{Procedure \S moveContents/}
Procedure \S moveContents/ has to move the contents of 
the current table to the next current table. All processes that 
have access to the table, may also participate in 
this migration. Indeed, they cannot yet use the new table (see invariants 
\S Ne1/ and \S Ne3/). 
We have to take care that delayed actions on the current 
table and the new table are carried out or abandoned correctly 
(see invariants \S Cu1/ and \S mE10/). 
Migration requires that every value in the current table 
be moved to a unique position in the new table (see invariant \S Ne19/).

Procedure \S moveContents/ uses a private variable \S toBeMoved/ 
that ranges over sets of locations. The procedure is given by
\begin {tab}
\>$\B proc /\S moveContents/
(\S from/, \S to/: \B pointer to /\S Hashtable/)=$\\
\>\>\B local/ $i:\S Nat/\;;\;b:\S Bool/\;;\; v:\S EValue/ \;;\;
\S toBeMoved/ : \B set/\ \B of/\ \S Nat/$ ;\\
\>\> $ \S toBeMoved/:= \{ 0,\ldots, \S from/.\T size/-1\}$ ;\\
110: \>\>$ \B while /\; \T currInd/ = \S index/\wedge
\S toBeMoved/ \ne \emptyset \;\B\ do / $\\
111: \>\>\>\B choose/ $i\in\S toBeMoved/$ ;\\
\>\>\> $ v:=\S from/.\T table/[i]$ ;\\
\>\>\> $ \B if /\; v=\B done/\; \B\ then/$\\
112: \>\>\>\>$\S toBeMoved/:=\S toBeMoved/-\{i\}$ ;\\
\>\>\>\B else/\\
114: \>\>\>\>\la~$b:=(v=\S from/.\T table/[i])$ ;\\
\>\>\>\>~~\B if/ $b$ \B then/ $\S from/.\T table/[i]:=
\S old/(\S val/(v))\;\B\ end / \ra$ ;\\
 \>\>\>\>\B if/ $b$ \B then/\\
116: \>\>\>\>\> $ \B if /\; \S val/(v)\ne \B null/\; 
\B\ then /\; \S moveElement/(\S val/(v),\S to/)\;\B\ end/$ ;\\
117: \>\>\>\>\>$\S from/.\T table/[i]:=\B done/ $ ;\\
118: \>\>\>\>\>$\S toBeMoved/:=\S toBeMoved/-\{i\}$ ;\\
 \>\>\B end end end /;\\
\>\B end/ .
\end {tab}

Note that the value is tagged as outdated before it is copied
(see invariant \S mC11/). 
After tagging, the value cannot be deleted or assigned until the 
migration has been completed. Tagging must be done atomically, 
since otherwise an interleaving deletion may be lost. When indeed 
the value has been copied to the new hash table, it becomes 
\B done/ in the old hash table in line 117. This has the effect 
that other processes need not wait for this process to complete 
procedure \S moveElement/, but can help with the migration of this 
value if needed.
 
Since the address is lost after being rewritten to \B done/, we 
had to introduce the shared auxiliary hash table \T Y/ to remember
its value for the proof of correctness. This could have been avoided 
by introducing a second tagging bit, say for ``very old''. 

The processes involved in the same migration should not 
use the same strategy for choosing $i$ in line 111, 
since it is advantageous that \S moveElement/ is called often with 
different values. They may exchange information: any of 
them may replace its set \S toBeMoved/ by the intersection 
of that set with the set \S toBeMoved/ of another one. 
We do not give a preferred strategy here, one can refer to 
algorithms for the {\it write-all} problem \cite{Jan00,KS97}.

\subsubsection{Procedure \S moveElement/}

The procedure \S moveElement/ moves a value 
to the new hash table. Note that the value is tagged as 
outdated in \S moveContents/ before \S moveElement/ is called. 
\begin {tab}
\> $\B proc /\S moveElement/(v:\S Value/\setminus\{\B null/\},
\S to/: \B pointer to /\S Hashtable/) = $ \\
\>\>\B local/ $a:\S Address/\;;\; k,m,n:\S Nat/\;;\; 
w:\S EValue/\;;\; b:\S Bool/$ ;\\
120: \>\>$n:=0\;;\; b:=\false\;;\; a :=\S ADR/(v)\;;\;
m:=\S to/.\T size/$ ;\\
\>\>\B repeat /\\
121: \>\>\>$k:=\; \S key/(a,m,n)\;;\; 
w:=\S to/.\T table/[k] $ ;\\
\>\>\>\B if/ $w=\B null/$ \B then/\\
123: \>\>\>\>\la~$b:=(\S to/.\T table/[k] = \B null/) $;\\
\>\>\>\>~~$ \B if /\; b\; \B\ then /\; \S to/.\T table/[k] :=v\ \B\ end /\ra$ ;\\
\>\>\>\B else/ $n\T ++/\;\B\ end/$ ;\\
125: \>\> $ \B until / \; b\Lor a=\S ADR/(w) \Lor
\T currInd/ \ne \S index/ $ ;\\
126: \>\> $ \B if /\; b\; \B\ then /\; \S to/.\T occ/\T ++/\;\B\ end/$ \\
\>\B end/ . 
\end {tab}

The value is only allowed to be inserted once in the new hash table 
(see invariant \S Ne19/), since otherwise the main property of open 
addressing would be violated. 
In total, four situations can occur in the procedure moveElement:
\begin{itemize}
\item {the current location $k$ contains a value with a different
address. The process increases $n$ to inspect the next location.}
\item {the current location $k$ contains a value with the same 
address. This means that the value has already been copied to the 
new hash table, the process therefore terminates.}
\item {the current location $k$ is an empty slot. The process inserts 
\S v/ and returns. If insertion fails, since another process filled 
the empty slot in between, the search is continued.}
\item {when \S index/ happens to differ from \T currInd/, 
the entire migration has been completed.}
\end{itemize}

While the current hash table pointer is not updated yet,
there exists at least one \B null/ entry in the new hash table 
(see invariants \S Ne8/, \S Ne22/ and \S Ne23/), 
hence the local variable $n$ in the 
procedure \S moveElement/ never goes beyond the size of the
 hash table (see invariants \S mE3/ and \S mE8/), 
and the termination is thus guaranteed.

\section {Correctness (Safety)} \label{safety}

In this section, we describe the proof of safety of the algorithm. 
The main aspects of safety are functional correctness, atomicity, and 
absence of memory loss. 
These aspects are formalized in eight invariants described in 
section \ref{safetymain}. To prove these invariants, we need many 
other invariants. These are listed in Appendix A. In section 
\ref{intuproof}, we sketch the verification of some of the invariants 
by informal means.  
In section \ref{pvs-model}, we describe how the theorem prover PVS is 
used in the verification. As exemplified in \ref{intuproof}, Appendix B 
gives the dependencies between the invariants. 

\B Notational Conventions./  
Recall that there are at most $P$ processes with process identifiers 
ranging from 1 up to $P$. We use $p$, $q$, $r$ to range over process 
identifiers, with a preference for $p$. 
Since the same program is executed by all processes, every private
variable name of a process $\ne p$ is extended with the suffix ``.'' +
``process identifier''. 
We do not do this for process $p$. So, e.g., the value of a private 
variable $x$ of process $q$ is denoted by $x.q$, but the value of
$x$ of process $p$ is just denoted by $x$. In particular, $pc.q$ is the 
program location of process $q$. It ranges over all integer labels used 
in the implementation. 

When local variables in different procedures have the same names, 
we add an abbreviation of the procedure name as a subscript to the name. 
We use the following abbreviations: \S fi/ for \S find/, \S del/ for 
\S delete/, \S ins/ for \S insert/, \S ass/ for \S assign/, \S gA/ for 
\S getAccess/, \S rA/ for \S releaseAccess/, \S nT/ for \S newTable/, 
\S mig/ for \S migrate/, \S ref/ for \S refresh/, \S mC/ for 
\S moveContents/, \S mE/ for \S moveElement/. 

In the implementation, there are several places where the same 
procedure is called, say \S getAccess/, \S releaseAccess/, etc. 
We introduce auxiliary private variables \S return/, local to such a
procedure, to hold the return location. We add a procedure subscript
to distinguish these variables according to the above convention. 

If $V$ is a set, $\sharp V$ denotes 
the number of elements of $V$. If $b$ is a boolean, then $\sharp b =0$ 
when $b$ is false, and $\sharp b = 1$ when $b$ is true. 
Unless explicitly defined otherwise, we always (implicitly) universally 
quantify over addresses $a$, values $v$, non-negative integer numbers 
$k$, $m$, and $n$, 
natural number $l$, processes $p$, $q$ and $r$. Indices $i$ and $j$ 
range over $[1,2P]$. We abbreviate \T H/(\T currInd/).\T size/ as 
\S curSize/.

In order to avoid using too many parentheses, we use the usual binding 
order for the operators. We give ``$\wedge$'' higher priority than
``$\vee$''. We use parentheses whenever necessary.

\subsection {Main properties}\label{safetymain}

We have proved the following three safety properties of the algorithm.
Firstly, the access procedures \S find/, \S delete/, \S insert/,
\S assign/, are functionally correct. Secondly they are executed 
atomically. The third safety property is absence of memory loss.

Functional correctness of \S find/, \S delete/, \S insert/ is the 
condition that the result of the implementation is the same as the 
result of the specification (fS), (dS), (iS). This is expressed by 
the required invariants:
\begin{tab} 
\S Co1:/ \>$ \S pc/=14 \Implies \S val/(r_{\it fi})=rS_{\it fi}$ \\
\S Co2:/ \>$ \S pc/\in \{25,26\} \Implies  
suc_{\it del}=sucS_{\it del}$ \\
\S Co3:/ \>$ \S pc/\in \{41,42\} \Implies  
suc_{\it ins}=sucS_{\it ins}$
\end{tab}

Note that functional correctness of \S assign/ holds trivially since it
 does not return a result.

According to the definition of atomicity in chapter 13 of \cite{Lyn96},
atomicity means that each execution of one of the access procedures
contains precisely one execution of the corresponding specifying action
(fS), (dS), (iS), (aS). We introduced the private auxiliary variables
\S cnt/ to count the number of times the specifying action is executed.
Therefore, atomicity is expressed by the invariants:
\begin{tab}
\S Cn1:/ \>$ \S pc/=14 \Implies \S cnt/_{\it fi}=1 $\\
\S Cn2:/ \>$ \S pc/\in\{25,26\} \Implies \S cnt/_{\it del}=1 $\\
\S Cn3:/ \>$ \S pc/\in\{41,42\} \Implies \S cnt/_{\it ins}=1 $\\
\S Cn4:/ \>$ \S pc/=57 \Implies \S cnt/_{\it ass}=1 $
\end{tab} 

We interpret absence of memory loss to mean that the number of 
allocated hash tables is bounded. More precisely, we prove that 
this number is bounded by $2P$. This is formalized in the invariant:
\begin{tab} 
\S No1:/ \>$ \sharp \{k \mid k < \T H/_-\T index/\Land 
\T Heap/(k)\neq\bot\}\leq 2P$ 
\end{tab}

An important safety property is that no
process accesses deallocated memory. Since most procedures perform
memory accesses, by means of pointers that are local variables, the
proof of this is based on a number of different invariants. Although,
this is not explicit in the specification, it has been checked 
because the theorem prover PVS does not allow access to
deallocated memory as this would violate type correctness conditions.

\subsection{Intuitive proof}\label{intuproof}

The eight correctness properties (invariants) mentioned above have 
been completely proved with the interactive proof checker of PVS. 
The use of PVS did not only take care of the delicate bookkeeping 
involved in the proof, it could also deal with many trivial cases 
automatically. At several occasions where PVS refused to let a proof 
be finished, we actually found a mistake and
 had to correct previous versions of this algorithm.

In order to give some feeling for the proof, we describe some proofs. 
For the complete mechanical proof, 
we refer the reader to \cite{HesMv}. Note that, for simplicity, we 
assume that all non-specific private variables in the proposed 
assertions belong to the general process $p$, and general process $q$ 
is an active process that tries to 
threaten some assertion ($p$ may equal $q$).\\

\noindent \B Proof/ of invariant \S Co1/ (as claimed in 
\ref{safetymain}). According to Appendix B, the 
stability of \S Co1/ follows from the invariants \S Ot3/, \S fi1/,
\S fi10/, which are given in Appendix A. 
Indeed, \S Ot3/ implies that no procedure returns to location 14. 
Therefore all return statements falsify the antecedent of \S Co1/ and 
thus preserve \S Co1/. Since $r_{\it fi}$ and $rS_{\it fi}$ are 
private variables to process $p$, \S Co1/ can only be violated by 
process $p$ itself (establishing $\S pc at 14/$) when $p$ executes 13 
with $r_{\it fi}=\B null/\Lor a_{\it fi}=\S ADR/(r_{\it fi})$. This 
condition is abbreviated as $\S Find/(r_{\it fi},a_{\it fi})$. 
Invariant \S fi10/ then implies that action 13 has the precondition 
$\S val/(r_{\it fi})=rS_{\it fi}$, so then it does not violate 
\S Co1/. In PVS, we used a slightly different definition of \S Find/, 
and we applied invariant \S fi1/ to exclude that $r_{\it fi}$ is 
\B done/ or \B del/, though invariant \S fi1/ is superfluous 
in this intuitive proof.\qed \\

\noindent \B Proof/ of invariant \S Ot3/. Since the procedures 
\S getAccess/, \S releaseAccess/, \S refresh/, \S newTable/ are 
called only at specific locations in the algorithm, it is easy to 
list the potential return addresses. Since the variables \S return/ 
are private to process $p$, they are not modified by other processes. 
Stability of \S Ot3/ follows from this. 
As we saw in the previous proof, \S Ot3/ is used to 
guarantee that no unexpected jumps occur.\qed \\

\noindent \B Proof/ of invariant \S fi10/. According to Appendix B, 
we only need to use \S fi9/ and \S Ot3/. Let us use the abbreviation 
$k = \S key/(a_{\it fi},l_{\it fi},n_{\it fi})$. Since $r_{\it fi}$ and 
$rS_{\it fi}$ are both private variables, they can only be modified by 
process $p$ when $p$ is executing statement 7. We split this situation 
into two cases
\begin{enumerate}
\item with precondition $\S Find/(h_{\it fi}.\T table/[k],a_{\it fi})$\\
After execution of statement 7,  $r_{\it fi}$ becomes $h_{\it fi}.
\T table/[k]$, 
and $rS_{\it fi}$ becomes $\T X/(a_{\it fi})$. By \S fi9/, we get 
$\S val/(r_{\it fi})=rS_{\it fi}$. Therefore the validity of \S fi10/ 
is preserved.
\item otherwise.\\
After execution of statement 7, $r_{\it fi}$ becomes 
$h_{\it fi}.\T table/[k]$, 
which then falsifies the antecedent of \S fi10/. \qed \\
\end{enumerate}

\noindent \B Proof/ of invariant \S fi9/. According to Appendix B, we 
proved that \S fi9/ follows 
from \S Ax2/, \S fi1/, \S fi3/, \S fi4/, \S fi5/, \S fi8/, 
\S Ha4/, \S He4/, \S Cu1/, \S Cu9/, \S Cu10/, and \S Cu11/.
We abbreviate $\S key/(a_{\it fi},l_{\it fi},n_{\it fi})$ as $k$.
We deduce $h_{\it fi}=\T H/(\S index/)$ from \S fi4/, 
$\T H/(\S index/)$ is not 
$\bot$ from \S He4/, and  $k$ is below $\T H/(\S index/).\T size/$ from 
\S Ax2/, \S fi4/ and \S fi3/. We split the proof into two cases:

\begin{enumerate}
\item $\S index/\neq\T currInd/$: By \S Ha4/, it follows that 
$\T H/(\S index/)\neq\T H/(\T currInd/)$. Hence from \S Cu1/, we obtain 
$h_{\it fi}.\T table/[k]=\B done/$, which falsifies the antecedent of 
\S fi9/. 
\item $\S index/=\T currInd/$: 
By premise $\S Find/(h_{\it fi}.\T table/[k],a_{\it fi})$, we know that 
$h_{\it fi}.\T table/[k]\neq\B done/$ because of \S fi1/. By \S Cu9/ 
and \S Cu10/, we obtain 
$\S val/(h_{\it fi}.\T table/[k])=\S val/(\T Y/[k])$. 
Hence it follows that $\S Find/(\T Y/[k],a_{\it fi})$. Using \S fi8/, 
we obtain 
\begin{tab}
\>$\forall m<n_{\it fi}:
\neg \S Find/(\T Y/[\S key/(a_{\it fi},\S curSize/,m)],a_{\it fi})$ 
\end{tab}
We get $n_{\it fi}$ is below $\S curSize/$ because of \S fi5/. 
By \S Cu11/, we conclude 
\begin{tab}
\>$\T X/(a_{\it fi})= \S val/(h_{\it fi}.\T table/[k])$ 
\end{tab}
\vspace{-5mm}\qed \\
\end{enumerate}

\subsection {The model in PVS} \label{pvs-model}

Our proof architecture (for one property) can be described as a 
dynamically growing tree in which each node is associated with an 
assertion. We start from a tree containing only one node, the proof 
goal, which characterizes some property of the system. We expand 
the tree by adding some new children via proper analysis of an 
unproved node (top-down approach, which requires a good understanding 
of the system). The validity of that unproved node is then reduced to 
the validity of its children and the validity of 
some less or equally deep nodes.

Normally, simple properties of the system are proved with appropriate 
precedence, and then used to help establish more complex ones. It is 
not a bad thing that some property that was taken for granted turns 
out to be not valid. Indeed, it may uncover a defect of the algorithm, 
but in any case it leads to new insights in it. 

We model the algorithm as a transition system \cite{ZA92}, 
which is described in the language of PVS in the following way. 
As usual in PVS, states are represented by a record with a number 
of fields:
{\small
\begin{tab}
\>  State : TYPE = $[\#$ \\
\>  \% \=global variables\\
\>       \>...\\
\>       \>busy : [ range(2*P) $\rightarrow$ nat ],\\
\>       \>prot : [ range(2*P) $\rightarrow$ nat ],\\
\>       \>...\\
\>  \% private variables:\\
\>       \>index : [ range(P) $\rightarrow$ range(2*P) ],\\
\>       \>...\\
\>       \>pc : [ range(P) $\rightarrow$ nat ], \% private program counters\\
\>       \>...\\
\>  \% local variables of procedures, also private to each process:\\
\>  \% find     \\
\>       \>a$_-$find : [ range(P) $\rightarrow$ Address ],\\
\>       \>r$_-$find : [ range(P) $\rightarrow$ EValue ],\\
\>       \>...\\
\>  \% getAccess   \\
\>       \>return$_-$getAccess : [ range(P) $\rightarrow$ nat ],\\
\>       \>...\\
\>  \#]
\end{tab}}
\noindent where \S range/(\R P/) stands for the range 
of integers from 1 to P.

Note that private variables are given with as argument a process 
identifier. Local variables are distinguished by adding their 
procedure's names as suffixes.

An action is a binary relation on states: it relates the state 
prior to the action to the state following the action. The system 
performed by a particular process is then specified by defining the 
precondition of each action as a predicate on the state and also the 
effect of each action in terms of a state transition. For example, 
line 5 of the algorithm is described in PVS as follows:
{\small
\begin{tab} 
\>\% corresponding to statement find5: $h:=\T H/[\S index/]$; $ n:=0 $;\\  
\>find5(\=i,s1,s2) : bool = \\
\>\>pc(s1)(i)=5 AND\\
\>\>s2 = s1 WITH [ \=(pc)(i) := 6,\\
\>\>\>(n$_-$find)(i) := 0,\\
\>\>\>(h$_-$find)(i) := H(s1)(index(s1)(i)) ]
\end{tab}}
\noindent
where $i$ is a process identifier, s1 is a pre-state, s2 
is a post-state. 

Since our algorithm is concurrent, the global transition relation is 
defined as the disjunction of all atomic actions.
{\small
\begin{tab} 
\>\% transition steps \\
\>step(\=i,s1,s2) : bool = \\
\>\> find5(i,s1,s2) or find6(i,s1,s2) or ... \\
\>\> delete15(i,s1,s2) or  delete16(i,s1,s2) or ...\\
\>\>...
\end{tab}}
\noindent
Stability for each invariant is proved by a PVS \S Theorem/ of the form:
{\small
\begin{tab} 
\>\% Theorem about the stability of invariant fi10 \\
\>IV$_-$fi10: THEOREM \\
\>\> forall (u,v : state, q : range(P) ) : \\
\>\>\>  step(q,u,v) AND fi10(u) AND fi9(u) AND ot3(u) \\
\>\>\>  =$>$ fi10(v)
\end{tab}}
\noindent
To ensure that all proposed invariants are stable, there is a 
global invariant \T INV/, which is the conjunction of all proposed invariants.
{\small
\begin{tab} 
\>\% global invariant \\
\>INV(s:state) : bool = \\
\>\> He3(s) and He4(s) and Cu1(s) and ... \\
\>\>...
\end{tab}
\begin{tab} 
\>\% Theorem about the stability of the global invariant INV \\
\>IV$_-$INV: THEOREM \\
\>\> forall (u,v : state, q : range(P) ) : \\
\>\>\>  step(q,u,v) AND INV(u)  =$>$ INV(v)
\end{tab}}
\noindent
We define \T Init/ as all possible initial states, for which all 
invariants must be valid.
{\small 
\begin{tab} 
\>\% initial state \\
\>Init: \{ s : state $\mid$ \\
\>\>\>(forall (p: range(P)):\\
\>\>\>\>  pc(s)(p)=0 and ...\\
\>\>\>\>  ...) and\\
\>\>\>(forall (a: Address):\\
\>\>\>\>  X(s)(a)=null) and \\
\>\>\>...\\
\>\>\>   \}
\end{tab}
\begin{tab} 
\>\% The initial condition can be satisfied by the global invariant 
INV \\
\>IV$_-$Init: THEOREM \\
\>\> INV(Init) 
\end{tab}}
\noindent
The PVS code contains ll
preconditions to imply well-definedness:
e.g. in \S find7/, the hash table must be non-NIL and $\ell$ 
must be its size. 
{\small 
\begin{tab} 
\>\% corresponding to statement find7\\  
\>find7(\=i,s1,s2) : bool = \\
\>\>  i?(Heap(s1)(h$_-$find(s1)(i))) and\\
\>\>  l$_-$find(s1)(i)=size(i$_-$(Heap(s1)(h$_-$find(s1)(i)))) and \\
\>\>  pc(s1)(i)=7 and\\
\>\>  ...
\end{tab}}

All preconditions are allowed, since we can prove lock-freedom in 
the following form. In every state $s1$ that satisfies the global 
invariant, every process $q$ can perform a step, i.e., there is a 
state $s2$ with $(s1,s2)\in\S step/$ and 
$\S pc/(s1,q)\ne\S pc/(s2,q)$. This is expressed in PVS by
{\small 
\begin{tab} 
\>\% theorem for lock-freedom\\
\>IV$_-$prog: THEOREM \\
\>\> forall (u: state, q: range(P) ) : \\
\>\>\>  INV(u) =$>$ exists (v: state): pc(u)(q) /= pc(v)(q) and step(q,u,v)
\end{tab}}

\section{Correctness (Progress)} \label{progress}

In this section, we prove that our algorithm is lock-free, and that 
it is wait-free for several subtasks.
Recall that an algorithm is called {\it lock-free} if always at least 
some process will finish its task in a 
finite number of steps, regardless of delays or failures by other 
processes. This means that no process can block the applications
of further operations to the data structure, although any particular 
operation need not terminate since a slow process can be passed 
infinitely often by faster processes.
We say that an operation is {\it wait-free} if any process involved
in that operation is guaranteed to complete it in a finite number of
its own steps, regardless of the (in)activity of other processes.

\subsection {The easy part of progress }

It is clear that \S releaseAccess/ is wait-free. It follows that 
the wait-freedom of \S migrate/ depends on wait-freedom of 
\S moveContents/. The loop of \S moveContents/ is clearly bounded. 
So, wait-freedom of \S moveContents/ depends on wait-freedom of 
\S moveElement/. It has been proved that $n$ 
is bounded by $m$ in \S moveElement/ (see invariants \S mE3/ and 
\S mE8/). Since, moreover, $\S to/.\T table/[k]\neq \B null/$ is 
stable, the loop of \S moveElement/ is also bounded. This concludes 
the sketch that \S migrate/ is wait-free. 

\subsection {Progress of newTable}

The main part of procedure \S newTable/ is wait-free. This can be 
shown informally, as follows. 
Since we can prove the condition $\T next/(\S index/)\neq 0$ is 
stable while process $p$ stays in the region $[77,84]$, once the 
condition $\T next/(\S index/)\neq 0$ 
holds, process $p$ will exit \S newTable/ in a few rounds.

Otherwise, we may assume that $p$ has precondition 
$\T next/(\S index/)= 0$ before executing line 78. By the invariant
\begin{tab}
\S Ne5:/ \>$ pc\in [1,58]\LOR pc\geq 62 \Land pc\neq 65 \Land 
\T next/(\S index/)=0$\\
\> $\implies\quad \S index/=\T currInd/$
\end{tab}
we get that $\S index/ = \T currInd/$ holds and 
$\T next/(\T currInd/)= 0$ from
the precondition. We define two sets of integers:
\begin{tab} 
\>\+ $ \S prSet1/(i) \IS 
\{r\mid \S index/.r=i \Land pc.r \notin \{0,59,60\}\}$\\

$\S prSet2/(i) \IS 
\{r\mid \S index/.r=i \Land pc.r \in \{104,105\} $\+\+\+\+\+ \\
$\lor~ i_{\it rA}.r=i \Land \S index/.r \neq i \Land pc.r 
\in [67,72] $\\
$\lor~ i_{\it nT}.r=i \Land pc.r \in [81,84]$\\
$\lor~ i_{\it mig}.r=i \Land pc.r \geq 97\;\}$
\end{tab}
and consider the sum 
$\sum_{i=1}^{2P}(\sharp (prSet1(i)) ~+~ \sharp (prSet2(i)))$.
While process $p$ is at line $78$, the sum cannot exceed $2P-1$
because there are only $P$ processes around and process $p$
contributes only once to the sum. It then follows from the pigeon 
hole principle that there exists $j\in [1,2P]$ such that 
$\sharp (prSet1(j)) ~+~ \sharp (prSet2(j))=0$
and $j\neq \S index/.p$. By the invariant
\begin{tab} 
\S pr1:/\> $ \T prot/[j] = \sharp (\S prSet1/(j)) + 
\sharp (\S prSet2/(j)) +
 \sharp (\T currInd/=j)$\\
\>\>\>\> $ + \sharp (\T next/(\T currInd/)=j)$ 
\end{tab} 
we can get that $\T prot/[j]=0$ because of 
$j\neq \S index/.p = \T currInd/$. 

While \T currInd/ is constant, no process can modify $\T prot/[j]$ for 
$j\neq \T currInd/$ infinitely often. Therefore, if process $p$ acts 
infinitely often and chooses its value $i$ in 78 by round robin, 
process $p$ exits the loop of \S newTable/ eventually. This shows 
that the main part of \S newTable/ is wait-free.

\subsection{The failure of wait-freedom}
Procedure \S getAccess/ is not wait-free. When the active clients 
keep changing the current index faster than the new client can 
observe it, the accessing client is doomed to starvation. In that case,
however, the other processes repeatedly succeed. It follows that 
\S getAccess/, \S refresh/, and \S newTable/ are lock-free. 

It may be possible to make a queue for the accessing clients which 
is emptied by a process in \S newTable/. The accessing clients 
must however also be able to enter autonomously. This would at 
least add another layer of complications. We therefore prefer 
to treat this failure of wait-freedom as a performance issue that
can be dealt with in practice by tuning the sizes of the hash tables. 

According to the invariants \S fi5/, \S de8/, \S in8/ and \S as6/, 
the primary procedures \S find/, \S delete/, \S insert/, 
\S assign/ are loops bounded by $n \leq h.\T size/$, and $n$ is only 
reset to 0 during migration. If $n$ is not reset to 0, it is 
incremented or stays constant. Indeed, the atomic \B if/ statements 
in 18b, 35b, and 50b have no \B else/ parts. In \S delete/ and 
\S assign/, it is therefore possible that $n$ stays constant without 
termination of the loop. Since 
\S assign/ can modify non-\B null/ elements of the table, it follows 
that \S delete/ and \S assign/ are not wait-free. 
This unbounded fruitless activity is possible only when \S assign/ 
actions of other processes repeatedly succeed. It follows that the 
primary procedures are lock-free. This concludes the 
argument that the system is lock-free. 

\section {Conclusions} \label{conclusions}

Lock-free shared data objects are inherently 
resilient to halting failures and permit maximum parallelism. We 
have presented a new practical, lock-free algorithm for 
concurrently accessible hash tables, which promises more robust 
performance and reliability than a conventional lock-based 
implementation. Moreover, the new algorithm is dynamic in the sense 
that it allows the hash table to grow and shrink as needed. 

The algorithm scales up linearly with the number of
processes, provided the function {\it key} and the selection
of $i$ in line 111 are defined well. This is confirmed by some
experiments where random values were stored, retrieved
and deleted from the hash table. These experiments indicated 
that $10^6$ insertions, deletions and finds per second and per 
processor are possible on an SGI powerchallenge with 250Mhz R12000 
processors. This figure should only be taken as a rough indicator, 
since the performance of parallel processing is very much influenced
by the machine architecture, the relative sizes of data structures
compared to sizes of caches, and even the scheduling of processes
on processors.

The correctness proof for our algorithm is noteworthy because of the 
extreme effort it took to finish it. Formal deduction by 
human-guided theorem proving can, in principle, verify any correct 
design, but doing so may require unreasonable amounts of effort, 
time, or skill. Though PVS provided great help for managing and 
reusing the proofs, we have to admit that the verification for our 
algorithm was very complicated due to the complexity of our 
algorithm. The total verification effort can roughly be estimated 
to consist of two man year excluding the effort in determining the 
algorithm and writing the documentation. The whole proof contains 
around 200 invariants. It takes an 1Ghz Pentium IV computer around 
two days to re-run an individual proof for one of the biggest 
invariants. Without suitable tool support like PVS, we even doubt 
if it would be possible to complete a 
reliable proof of such size and complexity.

It may well be possible to simplify the proof and reduce the number 
of invariants slightly, but we did not work on this. The complete 
version of the PVS specifications and the whole proof scripts can 
be found at \cite{HesMv}. Note that we simplified some definitions 
in the paper for the sake of presentation.\\

\appendix 
\section{Invariants} \label{Inv}
We present here all invariants whose validity has been verified 
by the theorem prover PVS.

\newenvironment{tab1}{\begin{tabbing}
MMMa\=aaa\=aaa\=aaa\=aaa\=aaa\=aaa\= \kill}{\end{tabbing}}
\medbreak Conventions. We abbreviate
\begin {tab1}
\>$\S Find/(\T r/,\T a/)\triangleq ~\T r/=\B null/\Lor \T a/=\S ADR/(\T r/)$ \\
\>$\S LeastFind/(a,n)\triangleq ~$\=$~(\forall m<n:\neg \S Find/(\T Y/[\S key/(a,\S curSize/,m)],a))$\\
\>\>$\land~\S Find/(\T Y/[\S key/(a,\S curSize/,n)],a))$\\
\> $\S LeastFind/(h,a,n)\triangleq ~$\=$~(\forall m<n:\neg \S Find/(h.\T table/[\S key/(a,h.\T size/,m)],a))$\\
\>\>$\land~\S Find/(h.\T table/[\S key/(a,h.\T size/,n)],a))$
\end {tab1}

\medbreak Axioms on functions \S key/ and \S ADR/
\begin{tab1} 
\S Ax1:/ \>$ v=\B null/\EQ \S ADR/(v)=\B 0/$ \\
\S Ax2:/ \>$ 0\leq \S key/(a,l,k) < l$\\
\S Ax3:/ \>$ 0\leq k < m < l ~\implies \quad \S key/(a,l,k)\neq \S key/(a,l,m)$
\end{tab1}

\medbreak Main correctness properties
\begin{tab1} 
\S Co1:/ \>$ pc=14 ~\implies \quad \S val/(r_{\it fi})=rS_{\it fi}$ \\
\S Co2:/ \>$ pc\in \{25,26\} ~\implies \quad  suc_{\it del}=sucS_{\it del}$ \\
\S Co3:/ \>$ pc\in \{41,42\} ~\implies \quad  suc_{\it ins}=sucS_{\it ins}$\\
\S Cn1:/ \>$ pc=14 ~\implies \quad cnt_{\it fi}=1 $\\
\S Cn2:/ \>$ pc\in\{25,26\} ~\implies \quad cnt_{\it del}=1 $\\
\S Cn3:/ \>$ pc\in\{41,42\} ~\implies \quad cnt_{\it ins}=1 $\\
\S Cn4:/ \>$ pc=57 ~\implies \quad cnt_{\it ass}=1 $
\end{tab1} 
The absence of memory loss is shown by
\begin{tab1}
\S No1:/ \>$ \sharp (nbSet1) \leq 2*P$ \\
\S No2:/ \>$ \sharp (nbSet1)=\sharp (nbSet2)$
\end{tab1}
where $nbSet1$ and $nbSet2$ are sets of integers, characterized by
\begin{tab1} 
$\S nbSet1/ \IS \{k \mid k < \T H/_-\T index/\Land \T Heap/(k)\neq\bot\}$\\

$\S nbSet2/ \IS 
\{i\mid \T H/(i)\neq 0 \lor (\exists r:pc.r=71 \land i_{\it rA}.r=i)\}$
\end{tab1}

\medbreak Further, we have the following definitions of sets of integers:

\begin{tab1}
$\S deSet1/ \IS \{k\mid k < \S curSize/\Land \T Y/[k]=\B del/\}$\\

$\S deSet2/ \IS \{r\mid \S index/.r=\T currInd/ \Land pc.r=25 \land 
suc_{\it del}.r\}$\\

$\S deSet3/\IS \{k\mid k < \T H/(\T next/(\T currInd/)).\T size/ $\\
\> $\Land \T H/(\T next/(\T currInd/)).\T table/[k]=\B del/\}$ \\

$\S ocSet1/\IS \{r\mid \S index/.r\neq \T currInd/$\\
\>\>$\lor\; pc.r\in [30,41] \Lor pc.r\in [46,57]$\\
\>\>$\lor\; pc.r \in [59,65]\Land return_{\it gA}.r\geq 30$ \\
\>\>$\lor\; $\=$pc.r \in [67,72]\land\; (return_{\it rA}.r=59\Land return_{\it gA}.r\geq 30$\\
\>\>\>\>$\lor\; return_{\it rA}.r=90\Land return_{\it ref}.r\geq 30)$ \\
\>\>$\lor\; (pc.r=90\Lor pc.r\in [104,105])\Land return_{\it ref}.r\geq 30\}$ \\

$\S ocSet2/ \IS \{r\mid pc.r \geq 125 \Land b_{\it mE}.r \Land 
to.r=\T H/(\T currInd/)\} $ \\

$\S ocSet3/ \IS 
\{r\mid \S index/.r=\T currInd/ \Land pc.r=41 \Land suc_{\it ins}.r$\\
\>\>$\Lor \S index/.r=\T currInd/ \Land pc.r=57\Land  r_{\it ass}.r=\B null/\}$\\

$\S ocSet4/\IS \{k\mid 
k < \S curSize/ \Land \S val/(\T Y/[k])\neq\B null/\}$\\

$\S ocSet5/\IS 
\{k\mid k <\T H/(\T next/(\T currInd/)).\T size/ $\\
\>\>$\Land \S val/(\T H/(\T next/(\T currInd/)).\T table/[k])\neq\B null/\}$\\

$\S ocSet6/\IS 
\{k\mid k < \T H/(\T next/(\T currInd/)).\T size/ $\\
\>\>$\Land \T H/(\T next/(\T currInd/)).\T table/[k]\neq\B null/\}$\\

$\S ocSet7/\IS \{r\mid  pc.r\geq 125\Land 
b_{\it mE}.r\Land to.r=\T H/(\T next/(\T currInd/))\}$ \\

$ \S prSet1/(i) \IS 
\{r\mid \S index/.r=i \Land pc.r \notin \{0,59,60\}\}$\\

$\S prSet2/(i) \IS 
\{r\mid \S index/.r=i \Land pc.r \in \{104,105\} $\\
\>\>$\lor~ i_{\it rA}.r=i \Land \S index/.r \neq i \Land pc.r 
\in [67,72] $\\
\>\>$\lor~ i_{\it nT}.r=i \Land pc.r \in [81,84]$\\
\>\>$\lor~ i_{\it mig}.r=i \Land pc.r \geq 97\}$\\

$\S prSet3/(i)\IS 
\{r\mid \S index/.r=i \Land pc.r \in [61,65]\cup [104,105] $\\
\>\>$\lor~ i_{\it rA}.r=i \Land pc.r=72 $\\
\>\>$\lor~ i_{\it nT}.r=i \Land pc.r \in [81,82]$\\
\>\>$\lor~ i_{\it mig}.r=i \Land pc.r \in [97,98]\}$\\

$\S prSet4/(i) \IS 
\{r\mid \S index/.r=i \Land pc.r \in [61,65] $\\
\>\>$\lor~ i_{\it mig}.r=i \Land pc.r \in [97,98]\}$ \\

$\S buSet1/(i) \IS 
\{r\mid \S index/.r=i$\\
\>\>$\land~ $\=$(pc.r \in [1,58]\cup (62,68]\Land pc.r\neq 65$\\
\>\>\>$\lor~ pc.r\in [69,72] \Land return_{\it rA}.r>59$\\
\>\>\>$\lor~ pc.r>72)\}$ \\

$\S buSet2/(i)\IS 
\{r\mid \S index/.r=i \Land pc.r=104$\\
\>\>$\lor~ i_{\it rA}.r=i \Land \S index/.r \neq i \Land pc.r \in [67,68] $\\
\>\>$\lor~ i_{\it nT}.r=i \Land pc.r \in [82,84]$\\
\>\>$\lor~ i_{\it mig}.r=i \Land pc.r \geq 100\}$
\end{tab1}

\medbreak We have the following invariants concerning the \T Heap/
\begin{tab1} 
\S He1:/ \>$ \T Heap/(0)=\bot$ \\

\S He2:/ \>$ \T H/(i)\neq 0\equiv \T Heap/(\T H/(i))\neq \bot$\\

\S He3:/\>$ \T Heap/(\T H/(\T currInd/))\neq \bot$ \\

\S He4:/\>$ pc \in [1,58] \Lor pc>65 \Land \neg(pc \in [67,72]\Land i_{\it rA}=\S index/)$\\
\>$\implies \quad \T Heap/(\T H/(\S index/))\neq \bot$ \\

\S He5:/\>$ \T Heap/(\T H/(i))\neq \bot ~\implies \quad\ \T H/(i).\T size/\geq P$\\

\S He6:/\>$ \T next/(\T currInd/)\neq 0~\implies \quad
\T Heap/(\T H/(\T next/(\T currInd/)))\neq\bot$ 
\end{tab1}

\medbreak Invariants concerning  hash table pointers
\begin{tab1}
\S Ha1:/ \>$ \T H/_-\T index/>0$ \\

\S Ha2:/ \>$ \T H/(i)<\T H/_-\T index/$ \\

\S Ha3:/ \>$ i\neq j\Land \T Heap/(\T H/(i))\neq \bot ~\implies \quad \T H/(i)\neq\T H/(j)$\\

\S Ha4:/ \>$ \S index/\neq \T currInd/~\implies \quad\ \T H/(\S index/)\neq \T H/(\T currInd/)$ \\
\end{tab1}

Invariants about counters for calling the specification.

\begin{tab1} 
\S Cn5:/ \>$ pc\in [6,7]~\implies \quad cnt$\=$_{\it fi}=0 $\\
\S Cn6:/ \>$ pc\in [8,13]$\\
\>$\lor~ pc\in [59,65] \land return_{\it gA}=10$\\
\>$\lor~ pc\in [67,72] \land ($\=$return_{\it rA}=59 \land return_{\it gA}=10$\\
\>\>$\lor~ return_{\it rA}=90 \land return_{\it ref}=10$\\
\>$\lor~ pc \geq 90 \Land return_{\it ref}=10$ \\
\>$\implies \quad cnt$\=$_{\it fi}=\sharp (r_{\it fi}=\B null/\lor a_{\it fi}=\S ADR/(r_{\it fi}))$ \\

\S Cn7:/ \>$ pc\in [16,21]\Land pc\neq 18 $\\
\>$\lor~ pc\in [59,65] \land return_{\it gA}=20$\\
\>$\lor~ pc\in [67,72] \land ($\=$return_{\it rA}=59 \land return_{\it gA}=20$\\
\>\>$\lor~ return_{\it rA}=90 \land return_{\it ref}=20$\\
\>$\lor~ pc \geq 90 \Land return_{\it ref}=20$ \\
\>$\implies \quad cnt_{\it del}= 0$\\

\S Cn8:/ \>$ pc=18 ~\implies \quad cnt_{\it del}=\sharp (r_{\it del}=\B null/)$\\

\S Cn9:/ \>$ pc\in [28,33]$\\
\>$\lor~ pc\in [59,65] \land return_{\it gA}=30$\\
\>$\lor~ pc\in [67,72] \land ($\=$return_{\it rA}=59 \land return_{\it gA}=30$\\
\>\>$\lor~ return_{\it rA}=77 \land return_{\it nT}=30$\\
\>\>$\lor~ return_{\it rA}=90 \land return_{\it ref}=30$\\
\>$\lor~ pc\in [77,84] \land return_{\it nT}=30 $\\
\>$\lor~ pc \geq 90 \Land return_{\it ref}=30$ \\
\>$\implies \quad cnt_{\it ins}= 0$ \\

\S Cn10:/ \>$ pc\in [35,37]$\\
\>$\lor~ pc\in [59,65] \land return_{\it gA}=36$\\
\>$\lor~ pc\in [67,72] \land ($\=$return_{\it rA}=59 \land return_{\it gA}=36$\\
\>\>$\lor~ return_{\it rA}=90 \land return_{\it ref}=36$\\  
\>$\lor~ pc \geq 90 \Land return_{\it ref}=36$ \\
\>$\implies \quad cnt_{\it ins}= \sharp (a_{\it ins}=\S ADR/(r_{\it ins})\lor suc_{\it ins})$\\

\S Cn11:/ \>$ pc\in [44,52]$\\
\>$\lor~ pc\in [59,65] \land return_{\it gA}\in \{46,51\}$\\
\>$\lor~ pc\in [67,72] \land ($\=$return_{\it rA}=59 \land return_{\it gA} \in \{46,51\}$\\
\>\>$\lor~ return_{\it rA}=77 \land return_{\it nT}=46$\\
\>\>$\lor~ return_{\it rA}=90 \land return_{\it ref} \in \{46,51\}$\\  
\>$\lor~ pc\in [77,84] \land return_{\it nT}=46 $\\
\>$\lor~ pc \geq 90 \Land return_{\it ref}\in \{46,51\}$ \\
\>$\implies \quad cnt_{\it ass}sign= 0$
\end{tab1}

\medbreak Invariants about old hash tables, current hash table and the auxiliary hash table \T Y/.
 Here, we universally quantify over all non-negative integers $n< \S curSize/$.
\begin{tab1}
\S Cu1:/ \>$ \T H/(\S index/)\neq \T H/(\T currInd/)\Land k < \T H/(\S index/).\T size/$\\
\>$\land~ (pc\in [1,58]\Lor pc>65 \Land 
\neg( pc\in[67,72]\Land i_{\it rA}=\S index/)$\\
\>$\implies\quad \T H/(\S index/).\T table/[k]=\B done/$ \\

\S Cu2:/ \>$\sharp(\{k\mid k < \S curSize/\Land \T Y/[k]\neq\B null/\})<\S curSize/$ \\

\S Cu3:/ \>$ \T H/(\T currInd/).\T bound/+2*P<\S curSize/$\\

\S Cu4:/ \>$ \T H/(\T currInd/).\T dels/+\sharp(deSet2)=\sharp(deSet1)$\\

\S Cu5:/ \>\S Cu5/ has been eliminated. The numbering has been kept, so as not to\\
\> endanger the consistency with Appendix B and the PVS script.\\

\S Cu6:/ \>$ \T H/(\T currInd/).\T occ/+\sharp (ocSet1)+\sharp (ocSet2)\leq\T H/(\T currInd/).\T bound/+2*P$\\

\S Cu7:/\>$\sharp(\{k\mid k < \S curSize/\Land\T Y/[k]\neq\B null/\}$ \\
\> $ = \T H/(\T currInd/).\T occ/+\sharp (ocSet2) + \sharp (ocSet3)$\\

\S Cu8:/ \>$\T next/(\T currInd/) = 0~\implies \quad
\neg\,\S oldp/(\T H/(\T currInd/).\T table/[n])$\\

\S Cu9:/ \>$\neg (\S oldp/(\T H/(\T currInd/).\T table/[n]))~\implies \quad 
\T H/(\T currInd/).\T table/[n]=\T Y/[n] $\\

\S Cu10:/ \>$\S oldp/(\T H/(\T currInd/).\T table/[n]) \Land \S val/(\T H/(\T currInd/).\T table/[n])\neq \B null/$\\
\>$\implies\ \S val/(\T H/(\T currInd/).\T table/[n])=\S val/(\T Y/[n])$ \\

\S Cu11:/ \>$\S LeastFind/(a,n)~\implies \quad \T X/(a)=\S val/(\T Y/[\S key/(a,\S curSize/,n)])$\\

\S Cu12:/ \>$\T X/(a)=\S val/(\T Y/[\S key/(a,\S curSize/,n)]) 
\neq \B null/~\implies \quad \S LeastFind/(a,n)$ \\

\S Cu13:/ \>$\T X/(a)=\S val/(\T Y/[\S key/(a,\S curSize/,n)])\neq \B null/
\Land n \ne m < \S curSize/$\\
\>$\implies \quad \S ADR/(\T Y/[\S key/(a,\S curSize/,m)])\neq a $ \\

\S Cu14:/ \>$\T X/(a)=\B null/\Land \S val/(\T Y/[\S key/(a,\S curSize/,n)])\neq \B null/$\\
\>$\implies \quad \S ADR/(\T Y/[\S key/(a,\S curSize/,n)])\neq a$ \\

\S Cu15:/ \>$\T X/(a)\neq\B null/$\\
\>$\implies\ \exists m < \S curSize/:\T X/(a)=\S val/(\T Y/[\S key/(a,\S curSize/,m)])$ \\

\S Cu16:/ \>$\exists (f:[$\=$\{m:0\leq m<\S curSize/)\land \S val/(\T Y/[m])\neq \B null/\}\rightarrow $\\
\>\>$\{v:v\neq\B null/\land (\exists k<\S curSize/:v=\S val/(\T Y/[k]))\}]):$\\
\>$f~ \T is bijective/$ 
\end{tab1}

\medbreak Invariants about \T next/ and $\T next/(\T currInd/)$:

\begin{tab1} 
\S Ne1:/ \>$ \T currInd/\neq\T next/(\T currInd/)$\\

\S Ne2:/ \>$ \T next/(\T currInd/)\neq 0~\implies \quad\T next/(\T next/(\T currInd/))=0$\\

\S Ne3:/ \>$ pc\in [1,59]\Lor pc\geq 62 \Land pc\neq 65~\implies \quad \S index/\neq\T next/(\T currInd/)$\\

\S Ne4:/ \>$ pc\in [1,58]\Lor pc\geq 62 \Land pc\neq 65~\implies \quad \S index/\neq\T next/(\S index/)$\\

\S Ne5:/ \>$ pc\in [1,58]\Lor pc\geq 62 \Land pc\neq 65 \Land \T next/(\S index/)=0$\\
\> $ \implies\quad \S index/=\T currInd/$\\

\S Ne6:/ \>$ \T next/(\T currInd/)\neq 0$\\
\>$\implies \quad $\=$ \sharp (\S ocSet6/)\leq \sharp(\{k\mid k < \S curSize/\Land \T Y/[k]\neq\B null/\}-\T H/(\T currInd/).\T dels/$\\
\>\>$-\sharp (deSet2)$\\

\S Ne7:/ \>$ \T next/(\T currInd/)\neq 0$\\
\>$\implies \quad $\=$\T H/(\T currInd/).\T bound/-\T H/(\T currInd/).\T dels/+2*P$\\
\>\>$\leq\T H/(\T next/(\T currInd/)).\T bound/$\\

\S Ne8:/ \>$ \T next/(\T currInd/)\neq 0$\\
\>$\implies \quad \T H/(\T next/(\T currInd/)).\T bound/+
2*P<\T H/(\T next/(\T currInd/)).\T size/$\\

\S Ne9:/ \>$ \T next/(\T currInd/)\neq 0 ~\implies \quad 
\T H/(\T next/(\T currInd/)).\T dels/=\sharp (\S deSet3/)$\\

\S Ne9a:/ \>$ \T next/(\T currInd/)\neq 0 ~\implies \quad 
\T H/(\T next/(\T currInd/)).\T dels/=0$ \\

\S Ne10:/ \>$ \T next/(\T currInd/)\neq 0 \Land k < h.\T size/~\implies \quad h.\T table/[k]\notin \{\B del/,\B done/\} $,\\
\>where $h=\T H/(\T next/(\T currInd/))$\\

\S Ne11:/ \>$\T next/(\T currInd/)\neq 0\Land 
k < \T H/(\T next/(\T currInd/)).\T size/ $\\
\> $\implies\quad \neg\S oldp/(\T H/(\T next/(\T currInd/)).\T table/[k])$\\

\S Ne12:/ \>$ k < \S curSize/\Land \T H/(\T currInd/).\T table/[k]=\B done/\Land  m< h.\T size/$\\
\>$\land\;\; \S LeastFind/(h,a,m) $\\
\> $ \implies\quad \T X/(a)=\S val/(h.\T table/[\S key/(a,h.\T size/,m)])$, \\
\> where $a=\S ADR/(\T Y/[k])$ and $h=\T H/(\T next/(\T currInd/)))$\\

\S Ne13:/ \>$k < \S curSize/\Land \T H/(\T currInd/).\T table/[k]=\B done/\Land m <h.\T size/$\\
\> $\land\;\; \T X/(a)=\S val/(h.\T table/[\S key/(a,h.\T size/,m)])
\neq \B null/$\\
\> $\implies\quad \S LeastFind/(h,a,m)$,\\
\> where $a=\S ADR/(\T Y/[k])$ and $h=\T H/(\T next/(\T currInd/))$\\

\S Ne14:/ \>$\T next/(\T currInd/)\neq 0 \Land a\neq \B 0/ \Land k < h.\T size/$\\
\>$\land\;\; \T X/(a)=\S val/(h.\T table/[\S key/(a,h.\T size/,k)])
\neq\B null/$\\
\>$\implies\quad \S LeastFind/(h,a,k)$,\\
\>where $h=\T H/(\T next/(\T currInd/))$\\

\S Ne15:/ \>$k < \S curSize/\Land \T H/(\T currInd/).\T table/[k]=\B done/\Land \T X/(a)\neq \B null/$\\
\>$ \land\;\; m < h.\T size/\Land \T X/(a)=\S val/(h.\T table/[\S key/(a,h.\T size/,m)])$\\
\>$ \land\;\; n < h.\T size/ \Land m \ne n $ \\
\> $\implies\quad \S ADR/(h.\T table/.[\S key/(a,h.\T size/,n)])\ne a$,\\
\>where $ a=\S ADR/(\T Y/[k])$ and $h=\T H/(\T next/(\T currInd/))$\\

\S Ne16:/ \>$k < \S curSize/\Land \T H/(\T currInd/).\T table/[k]=\B done/\Land \T X/(a)= \B null/ $\\
\>$\land\;\; m < h.\T size/$\\
\>$\implies\quad \S  val/(h.\T table/[\S key/(a,h.\T size/,m)]) = \B null/$\\
\>\>$\lor\;\; \S ADR/(h.\T table/[\S key/(a,h.\T size/,m)])\ne a $, \\
\>where $ a=\S ADR/(\T Y/[k])$ and $h=\T H/(\T next/(\T currInd/))$\\

\S Ne17:/ \>$ \T next/(\T currInd/)\neq 0\Land m < h.\T size/\Land a=\S ADR/(h.\T table/[m])\ne 0 $\\
\> $ \implies\quad \T X/(a)=\S val/(h.\T table/[m]) \neq \B null/$, \\
\>where $h=\T H/(\T next/(\T currInd/))$\\

\S Ne18:/ \>$ \T next/(\T currInd/)\neq 0\Land m < h.\T size/\Land a = \S ADR/(h.\T table/[m]) \ne 0 $\\
\> $ \implies\quad \exists n < \S curSize/: \S val/(\T Y/[n])=\S val/(h.\T table/[m]) $\\
\>\>$\land\;\; \S oldp/(\T H/(\T currInd/).\T table/[n])$, \\
\>where $h=\T H/(\T next/(\T currInd/))$\\

\S Ne19:/ \>$ \T next/(\T currInd/)\neq 0 \Land m < h.\T size/ \Land  m \ne n < h.\T size/$ \\
\> $ \land\;\; a = \S ADR/(h.\T table/[\S key/(a,h.\T size/,m)])\ne 0 $\\
\> $ \implies\quad \S ADR/(h.\T table/[\S key/(a,h.\T size/,n)])\ne a $,\\
\>where $h=\T H/(\T next/(\T currInd/))$\\

\S Ne20:/ \>$ k < \S curSize/\Land 
\T H/(\T currInd/).\T table/[k]=\B done/ \Land \T X/(a)\neq \B null/$\\
\>$\implies \quad \exists m < h.\T size/: 
\T X/(a)=\S val/(h.\T table/[\S key/(a,h.\T size/,m)])$,\\
\>where $a=\S ADR/(\T Y/ [k])$ and $ h=\T H/(\T next/(\T currInd/))$\\

\S Ne21:/ \>\S Ne21/ has been eliminated.\\

\S Ne22:/ \>$ \T next/(\T currInd/)\neq 0~$\\
\>$\implies \quad \sharp (\S ocSet6/)=\T H/(\T next/(\T currInd/)).\T occ/+\sharp (ocSet7)$\\

\S Ne23:/ \>$ \T next/(\T currInd/)\neq 0~$\\
\>$\implies \quad \T H/(\T next/(\T currInd/)).\T occ/\leq\T H/(\T next/(\T currInd/)).\T bound/$ \\

\S Ne24:/ \>$ \T next/(\T currInd/)\neq 0~\implies \quad 
\sharp (\S ocSet5/)\leq \sharp (\S ocSet4/)$ \\

\S Ne25:/ \>$~~~ \T next/(\T currInd/)\neq 0$\\
\>$\implies \quad \exists $\=$(f:[$\=$\{m:0\leq m<h.\T size/\land \S val/(h.\T table/[m])\neq \B null/\}\rightarrow $\\
\>\>$\{v:v\neq\B null/\land (\exists k<h.\T size/:v=\S val/(h.\T table/[k]))\}]):$\\
\>\>$f~ \T is bijective/$, \\
\>where $h=\T H/(\T next/(\T currInd/))$\\

\S Ne26:/ \>$~~~ \T next/(\T currInd/)\neq 0$\\
\>$\implies \quad \exists $\=$(f:[$\=$\{v:v\neq\B null/
\land (\exists m<h.\T size/:v=\S val/(h.\T table/[m]))\}\rightarrow $\\
\>\>\>$\{v:v\neq\B null/\land (\exists k:<\S curSize/:v=\S val/(\T Y/[k]))\}]):$\\
\>\>$f~ \T is injective/$, \\
\>where $h=\T H/(\T next/(\T currInd/))$\\

\S Ne27:/ \>$~~~ \T next/(\T currInd/)\neq 0 \Land (\exists n<h.\T size/: \S val/(h.\T table/[n])\neq \B null/)$\\
\>$\implies \quad \exists $\=$(f:[$\=$\{m:0\leq m<h.\T size/\land \S val/(h.\T table/[m]) \neq \B null/\} \rightarrow $\\
\>\>\>$\{k:0\leq k<\S curSize/\land \S val/(\T Y/[k])\neq \B null/\}])$\\
\>\>$f~ \T is injective/$, \\
\>where $h=\T H/(\T next/(\T currInd/))$
\end{tab1}

\medbreak Invariants concerning procedure \S find/ (5\dots 14)
\begin{tab1} 

\S fi1:/ \>$a_{\it fi}\neq \B 0/$ \\

\S fi2:/ \>$ pc\in \{6,11\}~\implies \quad n_{\it fi}=0$ \\

\S fi3:/ \>$ pc \in \{7,8,13\} ~\implies \quad l_{\it fi}=h_{\it fi}.\T size/$ \\

\S fi4:/ \>$ pc \in [6,13]\Land pc\neq 10~\implies \quad h_{\it fi}=\T H/(\S index/)$ \\

\S fi5:/ \>$ pc=7 \Land h_{\it fi}=\T H/(\T currInd/)~\implies \quad 
n_{\it fi} < \S curSize/$ \\

\S fi6:/ \>$ pc=8\Land h_{\it fi}=\T H/(\T currInd/)\Land\neg \S Find/(r_{\it fi},a_{\it fi})
\Land r_{\it fi}\neq\B done/$\\
\>$\implies\ \neg\ \S Find/(\T Y/[\S key/(a_{\it fi},\S curSize/,n_{\it fi})],
a_{\it fi})$ \\

\S fi7:/ \>$ pc=13 \Land h_{\it fi}=\T H/(\T currInd/)\Land \neg\S Find/(r_{\it fi},a_{\it fi})
\Land m < n_{\it fi}$ \\
\> $ \implies\quad \neg\S Find/(\T Y/[\S key/(a_{\it fi},\S curSize/,m)], a_{\it fi})$ \\

\S fi8:/ \>$ pc\in \{7,8\}\Land h_{\it fi}=\T H/(\T currInd/) \Land m < n_{\it fi}$\\
\> $ \implies\quad
\neg \S Find/(\T Y/[\S key/(a_{\it fi},\S curSize/,m)],a_{\it fi})$ \\

\S fi9:/ \>$ $\=$pc=7\Land\S Find/(t,a_{\it fi})~\implies \quad \T X/(a_{\it fi})=\S val/(t)$, \\
\>where $t=h_{\it fi}.\T table/[\S key/(a_{\it fi},l_{\it fi},n_{\it fi})]$\\

\S fi10:/ \>$  pc \notin (1,7] \Land \S Find/(r_{\it fi},a_{\it fi})~\implies \quad
\S val/(r_{\it fi})=rS_{\it fi} $ \\

\S fi11:/ \>$pc=8\Land \S oldp/(r_{\it fi})\Land \S index/=\T currInd/$\\
\>$\implies\ \T next/(\T currInd/)\neq 0$ 
\end{tab1}

\medbreak Invariants concerning procedure \S delete/ (15\dots 26)
\begin{tab1} 

\S de1:/ \>$a_{\it del}\neq \B 0/$ \\

\S de2:/ \>$ pc \in\{17,18\}~\implies \quad l_{\it del}=h_{\it del}.\T size/$ \\

\S de3:/ \>$ pc\in [16,25] \Land pc\neq 20~\implies \quad h_{\it del}=\T H/(\S index/)$ \\

\S de4:/ \>$ pc=18~\implies \quad k_{\it del}=\S key/(a_{\it del},l_{\it del},n_{\it del})$ \\

\S de5:/ \>$ pc\in \{16,17\} \Lor \S Deleting/ ~\implies \quad \neg suc_{\it del}$\\

\S de6:/ \> $ \S Deleting/ \Land sucS_{\it del}~\implies \quad r_{\it del}\neq \B null/$ \\

\S de7:/ \>$ pc=18\Land \neg\ \S oldp/(h_{\it del}.\T table/[k_{\it del}])~\implies \quad 
h_{\it del}=\T H/(\T currInd/)$ \\

\S de8:/ \>$ pc \in \{17,18\} \Land h_{\it del}=\T H/(\T currInd/)~\implies \quad
 n_{\it del}<\S curSize/$\\

\S de9:/ \>$pc=18\Land h_{\it del}=\T H/(\T currInd/)$\\
\>$\land~ (\S val/(r_{\it del})\neq\B null/\Lor r_{\it del}=\B del/)$\\
\>$\implies\ r\neq \B null/\Land (r=\B del/\Lor 
\S ADR/(r)=\S ADR/(r_{\it del}))$,\\
\>where $r=\T Y/[\S key/(a_{\it del},h_{\it del}.size,n_{\it del})]$\\

\S de10:/ \>$ pc\in\{17,18\}\Land h_{\it del}=\T H/(\T currInd/) \Land m < n_{\it del}) $\\
\> $ \implies\quad \neg \S Find/(\T Y/[\S key/(a_{\it del},\S curSize/,m)],a_{\it del})$\\

\S de11:/ \>$ pc\in \{17,18\} \Land \S Find/(t,a_{\it del})~\implies \quad 
\T X/(a_{\it del})=\S val/(t)$,\\
\>where $t=h_{\it del}.\T table/[\S key/(a_{\it del},l_{\it del},n_{\it del})]$\\

\S de12:/ \>$pc=18\Land \S oldp/(r_{\it del})\Land \S index/=\T currInd/$\\
\>$\implies\ \T next/(\T currInd/)\neq 0$ \\

\S de13:/ \>$ pc=18~\implies \quad k_{\it del}<\T H/(\S index/).\T size/$ \\
\end{tab1}

\medbreak\noindent
where $Deleting$ is characterized by
\begin{tab1}
\+ $\S Deleting/\EQ $\\
$pc\in [18,21]\Lor pc\in [59,65]\Land return_{\it gA}=20$\\
$\lor~ pc\in [67,72]\Land ($\=$return_{\it rA}=59 \land 
return_{\it gA}=20$\\
\>$\lor~ return_{\it rA}=90 \land return_{\it ref}=20)$\\
$\lor\ pc\geq 90 \Land  return_{\it ref}=20$
\end{tab1}

\medbreak Invariants concerning procedure \S insert/ (27\dots 52)

\begin{tab1}
\S in1:/ \>$a_{\it ins}=\S ADR/(v_{\it ins})\Land v_{\it ins}\neq \B null/$ \\

\S in2:/ \>$pc\in [32,35]~\implies \quad l_{\it ins}=h_{\it ins}.\T size/$ \\

\S in3:/ \>$pc\in [28,41]\Land pc\notin \{30,36\}~\implies \quad h_{\it ins}=\T H/(\S index/)$ \\

\S in4:/ \>$ pc\in \{33,35\}~\implies \quad 
k_{\it ins}=\S key/(a_{\it ins},l_{\it ins},n_{\it ins})$ \\

\S in5:/ \>$pc\in [32,33]\Lor \S Inserting/ ~\implies \quad \neg suc_{\it ins}$\\

\S in6:/ \> $\S Inserting/\Land sucS_{\it ins}~\implies \quad
\S ADR/(r_{\it ins})\neq a_{\it ins}$\\

\S in7:/ \>$ pc=35\Land \neg\ \S oldp/(h_{\it ins}.\T table/[k_{\it ins}])~\implies \quad 
h_{\it ins}=\T H/(\T currInd/)$ \\

\S in8:/ \>$ pc \in \{33,35\} \Land h_{\it ins}=\T H/(\T currInd/)~\implies \quad n_{\it ins}<\S curSize/$\\

\S in9:/ \>$pc=35\Land h_{\it ins}=\T H/(\T currInd/)$\\
\>$\land~ (\S val/(r_{\it ins})\neq\B null/\Lor r_{\it ins}=\B del/)$\\
\>$\implies\ r\neq \B null/\Land (r=\B del/\Lor \S ADR/(r)=\S ADR/(r_{\it ins}))$,\\
\>where $r=\T Y/[\S key/(a_{\it ins},h_{\it ins}.size,n_{\it ins})]$\\

\S in10:/ \>$pc\in \{32,33,35\}\Land h_{\it ins}=\T H/(\T currInd/) \Land m < n_{\it ins} $\\
\>$ \implies\quad \neg \S Find/(\T Y/[\S key/(a_{\it ins},\S curSize/,m)],a_{\it ins})$\\

\S in11:/ \>$pc\in \{ 33,35\}\Land \S Find/(t,a_{\it ins})~\implies \quad\ \T X/(a_{\it ins})=\S val/(t)$,\\
\>where $t=h_{\it ins}.\T table/[\S key/(a_{\it ins},l_{\it ins},n_{\it ins})]$ \\

\S in12:/ \>$pc=35\Land \S oldp/(r_{\it ins})\Land \S index/=\T currInd/$\\
\>$\implies\ \T next/(\T currInd/)\neq 0$ \\

\S in13:/ \>$ pc=35~\implies \quad k_{\it ins}<\T H/(\S index/).\T size/$ \\
\end{tab1}
where $Inserting$ is characterized by
\begin{tab1}
\+ $\S Inserting/ \EQ $\\
$pc\in [35,37]\Lor pc\in [59,65]\Land return_{\it gA}=36$\\
$\lor~ pc\in [67,72]\Land ($\=$return_{\it rA}=59 \land 
return_{\it gA}=36$\\
\>$\lor~ return_{\it rA}=90 \land return_{\it ref}=36)$\\
$\lor~ pc\geq 90 \Land  return_{\it ref}=36$
\end{tab1}

\medbreak Invariants concerning procedure \S assign/ (43\dots 57)

\begin{tab1}
\S as1:/ \>$ a_{\it ass}=\S ADR/(v_{\it ass})\Land v_{\it ass}\neq \B null/$ \\

\S as2:/ \>$pc\in [48,50]~\implies \quad l_{\it ass}=h_{\it ass}.\T size/$ \\

\S as3:/ \>$pc\in [44,57]\Land pc\notin \{46,51\}~\implies \quad h_{\it ass}=\T H/(\S index/)$ \\

\S as4:/ \>$ pc\in \{49,50\}~\implies \quad 
k_{\it ass}=\S key/(a_{\it ass},l_{\it ass},n_{\it ass})$\\

\S as5:/ \>$ pc=50\Land \neg\ \S oldp/(h_{\it ass}.\T table/[k_{\it ass}])~\implies \quad h_{\it ass}=\T H/(\T currInd/)$\\

\S as6:/ \>$ pc = 50 \Land h_{\it ass}=\T H/(\T currInd/)~\implies \quad n_{\it ass}<\S curSize/$ \\

\S as7:/ \>$pc=50\Land h_{\it ass}=\T H/(\T currInd/)$\\
\>$\land~ (\S val/(r_{\it ass})\neq\B null/\Lor r_{\it ass}=\B del/)$\\
\>$\implies\ r\neq \B null/\Land (r=\B del/\Lor \S ADR/(r)=\S ADR/(r_{\it ass}))$,\\
\>where $r=\T Y/[\S key/(a_{\it ass},h_{\it ass}.size,n_{\it ass})]$ \\

\S as8:/ \>$pc\in \{48,49,50\}\Land h_{\it ass}=\T H/(\T currInd/) \Land m < n_{\it ass} $\\
\> $ \implies\quad \neg \S Find/(\T Y/[\S key/(a_{\it ass},\S curSize/,m)],a_{\it ass})$\\

\S as9:/ \>$pc=50 \Land \S Find/(t,a_{\it ass})~\implies \quad \T X/(a_{\it ass})=\S val/(t)$,\\
\>where $t=h_{\it ass}.\T table/[\S key/(a_{\it ass},l_{\it ass},n_{\it ass})]$\\

\S as10:/ \>$pc=50\Land \S oldp/(r_{\it ass}sign)\Land \S index/=\T currInd/$\\
\>$\implies\ \T next/(\T currInd/)\neq 0$ \\

\S as11:/ \>$ pc=50~\implies \quad k_{\it ass}<\T H/(\S index/).\T size/$ \\
\end{tab1}

\medbreak Invariants concerning procedure \S releaseAccess/ (67\dots 72)

\begin{tab1}

\S rA1:/ \>$ h_{\it rA}<\T H/_-\T index/$\\

\S rA2:/ \>$ pc\in [70,71]~\implies \quad h_{\it rA}\neq 0$ \\

\S rA3:/ \>$pc=71 ~\implies \quad \T Heap/(h_{\it rA})\neq\bot$ \\

\S rA4:/ \>$ pc=71~\implies \quad \T H/(i_{\it rA})=0$ \\

\S rA5:/ \>$ pc=71~\implies \quad h_{\it rA}\ne\T H/(i)$ \\

\S rA6:/ \>$pc=70~\implies \quad\T H/(i_{\it rA})\neq\T H/(\T currInd/)$ \\

\S rA7:/ \>$pc=70$\\
\>$\land\ (pc.r \in [1,58]\Lor pc.r>65\land 
\neg (pc.r \in [67,72]\land i_{\it rA}.r=\S index/.r))$\\
\>$\implies\ \T H/(i_{\it rA})\neq \T H/(\S index/.r)$\\

\S rA8:/ \>$pc=70~\implies \quad i_{\it rA}\neq\T next/(\T currInd/)$ \\

\S rA9:/ \>$pc\in [68,72]\land\ (h_{\it rA}=0\Lor h_{\it rA}\neq \T H/(i_{\it rA}))$\\
\> $\implies\ \T H/(i_{\it rA})=0$ \\

\S rA10:/ \>$pc\in [67,72]\Land return_{\it rA}\in \{0,59\}~\implies\ i_{\it rA}=\S index/$\\

\S rA11:/ \>$pc\in [67,72]\Land return_{\it rA}\in \{77,90\}~\implies\ i_{\it rA}\neq \S index/$ \\

\S rA12:/ \>$pc\in [67,72]\Land return_{\it rA}=77~\implies \quad\T next/(\S index/)\neq 0$ \\

\S rA13:/ \>$pc=71 \Land pc.r=71 \Land p\neq r ~\implies \quad h_{\it rA}\neq h_{\it rA}.r$ \\

\S rA14:/ \>$pc=71 \Land pc.r=71 \Land p\neq r ~\implies \quad i_{\it rA}\neq i_{\it rA}.r$
\end{tab1}

\medbreak Invariants concerning procedure \S newTable/ (77\dots 84)

\begin{tab1}

\S nT1:/ \>$  pc \in [81,82] ~\implies \quad \T Heap/(\T H/(i_{\it nT}))=\bot$ \\

\S nT2:/ \>$ pc \in [83,84]~\implies \quad \T Heap/(\T H/(i_{\it nT}))\neq \bot$ \\

\S nT3:/ \>$ pc=84 ~\implies \quad\T next/(i_{\it nT})=0$ \\

\S nT4:/ \>$ pc \in [83,84] ~\implies \quad \T H/(i_{\it nT}).\T dels/=0$ \\

\S nT5:/ \>$  pc \in [83,84] ~\implies \quad \T H/(i_{\it nT}).\T occ/=0$ \\

\S nT6:/ \>$  pc \in [83,84] ~\implies \quad 
\T H/(i_{\it nT}).\T bound/+2*P< \T H/(i_{\it nT}).\T size/$ \\

\S nT7:/ \>$  pc \in [83,84] \Land \S index/=\T currInd/$\\
\>$\implies \quad \T H/(\T currInd/).\T bound/-\T H/(\T currInd/).\T dels/+2*P< \T H/(i_{\it nT}).\T bound/$ \\

\S nT8:/ \>$ pc \in [83,84]\Land k < \T H/(i_{\it nT}).\T size/
~\implies \quad \T H/(i_{\it nT}).\T table/[k]=\B null/$ \\

\S nT9:/ \>$ pc \in [81,84]~\implies \quad i_{\it nT}\neq \T currInd/$ \\

\S nT10:/ \>$ pc \in [81,84]\Land (pc.r\in [1,58]\Lor pc.r\geq 62 \Land pc.r\neq 65) $\\
\>$\implies\quad i_{\it nT}\ne \S index/.r$\\

\S nT11:/ \>$ pc\in [81,84]~\implies \quad i_{\it nT}\neq \T next/(\T currInd/)$\\

\S nT12:/ \>$ pc \in [81,84]~\implies \quad\T H/(i_{\it nT})\neq \T H/(\T currInd/)$\\

\S nT13:/ \>$ pc \in [81,84]$\\
\>$\land\ (pc.r \in [1,58]\Lor pc.r>65\land \neg (pc.r \in [67,72]\land i_{\it rA}.r=\S index/.r))$\\
\>$\implies\quad \T H/(i_{\it nT})\ne \T H/(\S index/.r)$ \\

\S nT14:/ \>$ pc\in [81,84]\Land pc.r\in [67,72]~\implies \quad 
i_{\it nT}\ne i_{\it rA}.r$\\

\S nT15:/ \>$ pc \in [83,84]\Land pc.r\in [67,72]~\implies \quad 
\T H/(i_{\it nT})\ne \T H/(i_{\it rA}.r)$ \\

\S nT16:/ \>$ pc\in [81,84]\Land pc.r\in [81,84]\Land p \ne r
~\implies \quad i_{\it nT}\ne i_{\it nT}.r$ \\

\S nT17:/ \>$ pc\in [81,84]\Land pc.r \in [95,99]\Land \S index/.r=\T currInd/$\\
\> $\implies\quad i_{\it nT}\ne i_{\it mig}.r$\\

\S nT18:/ \>$ pc\in [81,84]\Land pc.r\geq 99~\implies \quad  i_{\it nT}\ne i_{\it mig}.r$
\end{tab1}

\medbreak Invariants concerning procedure \S migrate/ (94\dots 105)

\begin{tab1}

\S mi1:/ \>$ pc=98 \Lor pc\in \{104,105\} ~\implies \quad \S index/\neq \T currInd/$\\

\S mi2:/ \>$ pc\geq 95~\implies \quad i_{\it mig}\neq \S index/$ \\

\S mi3:/ \>$ pc=94~\implies \quad \T next/(\S index/)>0$\\

\S mi4:/ \>$ pc\geq 95~\implies \quad i_{\it mig}\neq 0$\\

\S mi5:/ \>$ pc \geq 95~\implies \quad i_{\it mig}=\T next/(\S index/)$ \\

\S mi6:/ \>$ pc.r=70$\\
\>$\land~ (pc\in [95,102) \Land \S index/=\T currInd/ \Lor pc \in [102,103] \Lor pc\geq 110)$\\
\>$\implies \quad i_{\it rA}.r\neq i_{\it mig}$ \\

\S mi7:/ \>$ pc \in [95,97]\Land \S index/=\T currInd/\Lor pc\geq 99$\\
\>$\implies \quad i_{\it mig}\neq \T next/(i_{\it mig})$\\

\S mi8:/ \>$ (pc \in [95,97]\Lor pc \in [99,103] \Lor pc\geq 110)\Land \S index/=\T currInd/$\\
\>$\implies \quad \T next/(i_{\it mig})=0$\\

\S mi9:/ \>$(pc \in [95,103]\Lor pc\geq 110)\Land \S index/=\T currInd/$ \\
\>\>$\implies \quad \T H/(i_{\it mig})\neq \T H/(\T currInd/)$\\

\S mi10:/ \>$(pc \in [95,103]\Lor pc\geq 110)\Land \S index/=\T currInd/ $ \\
\> $ \land\;\; (pc.r \in [1,58]\Lor pc.r\geq 62\Land pc.r\neq 65)$\\
\>$\implies\quad \T H/(i_{\it mig}) \neq \T H/(\S index/.r) $ \\

\S mi11:/ \>$ pc=101 \Land \S index/=\T currInd/ \Lor pc=102$\\
\>$\implies \quad h_{\it mig}=\T H/(i_{\it mig})$\\

\S mi12:/ \>$pc\geq 95\Land \S index/=\T currInd/ \Lor pc\in \{102,103\}\Lor pc\geq 110$\\
\>$\implies \quad\ \T Heap/(\T H/(i_{\it mig}))\neq \bot$ \\

\S mi13:/ \>$pc=103\Land \S index/=\T currInd/\Land k < \S curSize/~$\\
\>$\implies \quad \T H/(\S index/).\T table/[k]=\B done/$\\

\S mi14:/ \>$$\=$pc=103\Land \S index/=\T currInd/\Land n < \T H/(i_{\it mig}).\T size/$\\
\>\>$\land\quad \S LeastFind/(\T H/(i_{\it mig}),a,n)$\\
\>\>$\implies\quad \T X/(a)=\S val/(\T H/(i_{\it mig})[\S key/(a,\T H/(i_{\it mig}).\T size/,n)])$ \\

\S mi15:/ \>$$\=$pc=103\Land \S index/=\T currInd/\Land n < \T H/(i_{\it mig}).\T size/$\\
\>\>$\land\quad \T X/(a)=
\S val/(\T H/(i_{\it mig}).\T table/[\S key/(a,\T H/(i_{\it mig}).\T size/,n)]
\neq\B null/$\\
\>\>$\implies\quad \S LeastFind/(\T H/(i_{\it mig}),a,n)$ \\

\S mi16:/ \>$pc=103\Land \S index/=\T currInd/\Land k < \T H/(i_{\it mig}).\T size/$\\
\>$ \implies\quad \neg \S oldp/(\T H/(i_{\it mig}).\T table/[k])$\\

\S mi17:/ \>$pc=103\Land \S index/=\T currInd/ \Land \T X/(a)\neq \B null/\Land 
k < h.\T size/$\\
\> $ \land\;\; \T X/(a)=\S val/(h.\T table/[\S key/(a,h.\T size/,k)])
\Land k \ne n < h.\T size/$\\
\> $\implies\quad \S ADR/(h.\T table/.[\S key/(a,h.\T size/,n)])\ne a $,\\
\>where $ h=\T H/(i_{\it mig})$ \\

\S mi18:/ \>$pc=103\Land \S index/=\T currInd/ \Land \T X/(a)=\B null/\Land k < h.\T size/$\\
\>$ \implies\quad \S val/(h.\T table/[\S key/(a,h.\T size/,k)]) = \B null/$\\
\>\>$\Lor~ \S ADR/(h.\T table/[\S key/(a,h.\T size/,k)])\ne a $, \\
\>where $ h=\T H/(i_{\it mig})$\\

\S mi19:/ \>$ pc=103\Land \S index/=\T currInd/\Land \T X/(a)\neq \B null /$\\
\> $ \implies\quad \exists m < h.\T size/: 
\T X/(a)=\S val/(h.\T table/[\S key/(a,h.\T size/,m)]$, \\
\>where $h=\T H/(i_{\it mig})$\\

\S mi20:/ \>$ $\=$pc=117 \Land \T X/(a)\neq \B null/
\Land \S val/(\T H/(\S index/).\T table/[i_{\it mC}])\neq \B null/$\\
\>$\lor~ pc\geq 126\Land \T X/(a)\neq \B null/\Land \S index/=\T currInd/$\\
\>$\lor~ pc=125 \Land \T X/(a)\neq \B null/\Land \S index/=\T currInd/$\\
\>\>$ \Land $\=$(b_{\it mE} \Lor \S val/(w_{\it mE})\neq \B null/$\\
\>\>\>$\Land a_{\it mE}=\S ADR/(w_{\it mE}))$ \\
\>$\implies\ \exists m < h.\T size/: 
\T X/(a)=\S val/(h.\T table/[\S key/(a,h.\T size/,m)])$,\\
\>where $a=\S ADR/(\T Y/[i_{\it mC}])$ and $h=\T H/(\T next/(\T currInd/))$
\end{tab1}

\medbreak Invariants concerning procedure \S moveContents/ (110\dots 118):
\begin{tab1}

\S mC1:/ \>$ pc=103 \Lor pc \geq 110~\implies \quad to=\T H/(i_{\it mig})$\\

\S mC2:/ \>$ pc\geq 110~\implies \quad from=\T H/(\S index/)$\\

\S mC3:/ \>$ pc>102\Land m \in toBeMoved~\implies \quad m<\T H/(\S index/).\T size/$\\

\S mC4:/ \>$ pc=111~\implies \quad \exists m < from.\T size/:m\in toBeMoved$ \\

\S mC5:/ \>$ pc\geq 114\Land pc\neq 118~\implies \quad v_{\it mC}\neq\B done/$ \\

\S mC6:/ \>$ pc\geq 114~\implies \quad i_{\it mC}<\T H/(\S index/).\T size/$ \\

\S mC7:/ \>$ pc=118~\implies \quad \T H/(\S index/).\T table/[i_{\it mC}]=\B done/$ \\

\S mC8:/ \>$pc\geq 110 \Land k < \T H/(\S index/).\T size/\Land 
k\notin toBeMoved$\\
\>$\implies \quad \T H/(\S index/).\T table/ [k]=\B done/$\\

\S mC9:/ \>$ pc \geq 110\Land \S index/=\T currInd/\Land toBeMoved=\emptyset $\\
\>$\land\;\; k < \T H/(\S index/).\T size/$\\
\> $\implies \quad \T H/(\S index/).\T table/[k]=\B done/$ \\

\S mC10:/ \>$ pc\geq 116\Land \S val/(v_{\it mC})\neq\B null/$\\
\>$\land~ \T H/(\S index/).\T table/[i_{\it mC}]=\B done/$ \\
\>$\implies \quad \T H/(i_{\it mig}).\T table/[\S key/(a,\T H/(i_{\it mig}).\T size/,0)]\neq \B null/$, \\
\>where $a=\S ADR/(v_{\it mC})$ \\

\S mC11:/ \>$ pc\geq 116 \Land\T H/(\S index/).\T table/[i_{\it mC}]\neq \B done/$ \\
\>$\implies\quad \S val/(v_{\it mC})=\S val/(\T H/(\S index/).\T table/[i_{\it mC}])$\\
\>$\Land~ \S oldp/(\T H/(\S index/).\T table/[i_{\it mC}])$\\

\S mC12:/ \>$ pc\geq 116\Land \S index/=\T currInd/\Land\S val/(v_{\it mC})\neq\B null/$ \\
\>$\implies \quad \S val/(v_{\it mC})=\S val/(\T Y/[i_{\it mC}])$ 
\end{tab1}

\medbreak Invariants concerning procedure \S moveElement/ (120\dots 126):

\begin{tab1} 
\S mE1:/ \>$ pc\geq 120~\implies \quad \S val/(v_{\it mC})=v_{\it mE}$ \\

\S mE2:/ \>$ pc \geq 120 ~\implies \quad v_{\it mE}\neq \B null/$\\

\S mE3:/ \>$ pc \geq 120 ~\implies \quad to=\T H/(i_{\it mig})$\\

\S mE4:/ \>$ pc \geq 121~\implies \quad a_{\it mE}=\S ADR/(v_{\it mC})$\\

\S mE5:/ \>$ pc\geq 121~\implies \quad m_{\it mE}=to.\T size/$\\

\S mE6:/ \>$ pc\in \{121,123\}~\implies \quad \neg b_{\it mE}$\\

\S mE7:/ \>$ pc=123 ~\implies \quad k_{\it mE}=\S key/(a_{\it mE},to.\T size/,n_{\it mE})$\\

\S mE8:/ \>$ pc\geq 123~\implies \quad k_{\it mE}<\T H/(i_{\it mig}).\T size/$\\

\S mE9:/ \>$ pc=120 $\\
\>$\land~ to.\T table/[\S key/(\S ADR/(v_{\it mE}),to.\T size/,0)]=\B null/$\\
\>$\implies \quad \S index/=\T currInd/$\\

\S mE10:/ \>$ pc \in \{121,123\}$\\
\>$\land~ to.\T table/[\S key/(a_{\it mE},to.\T size/,n_{\it mE})]=\B null/$\\
\>$\implies \quad \S index/=\T currInd/$\\

\S mE11:/ \>$ pc \in \{121,123\}\Land pc.r=103 $\\
\>$\land~ to.\T table/[\S key/(a_{\it mE},to.\T size/,n_{\it mE})]=\B null/$\\
\>$\implies\quad \S index/.r\ne \T currInd/$\\

\S mE12:/ \>$pc\in \{121,123\}\Land \T next/(\T currInd/)\neq 0 \Land to=\T H/(\T next/(\T currInd/))$\\
\>$\implies \quad n_{\it mE}<\T H/(\T next/(\T currInd/)).\T size/$ \\

\S mE13:/ \>$ pc \in \{123,125\}\Land w_{\it mE}\neq\B null/$\\
\>$\implies\quad \S ADR/(w_{\it mE})=\S ADR/(to.\T table/[k_{\it mE}])$\\
\>$~\Lor~ to.\T table/[k_{\it mE}]\in \{\B del/,\B done/\}$\\

\S mE14:/ \>$ pc \geq 123 \Land w_{\it mE}\neq\B null/$\\
\>$\implies \quad \T H/(i_{\it mig}).\T table/[k_{\it mE}]\neq\B null/$\\

\S mE15:/ \>$ pc=117 \Land \S val/(v_{\it mC})\neq \B null/$\\
\>$\lor~ pc\in \{121,123\}\Land n_{\it mE}>0$\\
\>$\lor~ pc=125$\\
\>$\implies \quad h.\T table/[\S key/(\S ADR/(v_{\it mC}),h.\T size/,0)]
\neq\B null/$,\\
\>where $h=\T H/(i_{\it mig})$\\

\S mE16:/ \>$ pc \in \{121,123\}$\\
\>$\lor~ ($\=$pc=125 \Land \neg b_{\it mE}$\\
\>\>$\land~ (\S val/(w_{\it mE})=\B null/ \Lor a_{\it mE}\neq \S ADR/(w_{\it mE})))$\\
\>$\implies \quad \forall m<n_{\it mE}:$\\
\>\>~~$\neg \S Find/(to.\T table/[\S key/(a_{\it mE},to.\T size/,m)],a_{\it mE})$
\end{tab1}

\medbreak Invariants about the integer array \T prot/.
\begin{tab1} 

\S pr1:/ \> $ \T prot/[i] ~=~ $\=$ \sharp (\S prSet1/(i)) ~+~ \sharp (\S prSet2/(i)) ~+~ \sharp (\T currInd/=i)$\\
\>\>$ +~ \sharp (\T next/(\T currInd/)=i)$ \\

\S pr2:/ \>$ \T prot/[\T currInd/]>0$ \\

\S pr3:/ \>$ pc \in [1,58]\Lor pc\geq 62 \Land pc\neq 65~\implies \quad \T prot/[\S index/]>0$ \\

\S pr4:/ \>$ \T next/(\T currInd/)\neq 0~\implies \quad \T prot/[\T next/(\T currInd/)]>0$\\

\S pr5:/ \>$ \T prot/[i]=0~\implies \quad \T Heap/(\T H/[i])=\bot$ \\

\S pr6:/ \>$ \T prot/[i]\leq\sharp (\S prSet3/(i)) \Land \T busy/[i]=0~\implies \quad
\T Heap/(\T H/[i])=\bot$ \\

\S pr7:/ \>$ pc \in [67,72]~\implies \quad \T prot/[i_{\it rA}]>0$ \\

\S pr8:/ \>$ pc\in [81,84]~\implies \quad \T prot/[i_{\it nT}]>0$\\

\S pr9:/ \>$ pc\geq 97~\implies \quad \T prot/[i_{\it mig}]>0$ \\

\S pr10:/ \>$ pc\in [81,82]~\implies \quad \T prot/[i_{\it nT}]=\sharp (prSet4(i_{\it nT}))+1$
\end{tab1}

\medbreak Invariants about the integer array \T busy/.

\begin{tab1} 
\S bu1:/ \> $ \T busy/[i] ~=~ $\=$\sharp (buSet1(i)) ~+~ \sharp (buSet2(i)) ~+~ \sharp (\T currInd/=i)$\\
\>\>$ +~ \sharp (\T next/(\T currInd/)=i) $\\

\S bu2:/ \>$ \T busy/[\T currInd/]>0$\\

\S bu3:/ \>$ pc \in [1,58]$\\
\>$\lor~ pc>65 \Land \neg(i_{\it rA}=\S index/ \Land pc \in [67,72])$\\
\>$\implies \quad \T busy/[\S index/]>0$\\

\S bu4:/ \>$ \T next/(\T currInd/)\neq 0~\implies \quad \T busy/[\T next/(\T currInd/)]>0$\\

\S bu5:/ \>$ pc=81~\implies \quad \T busy/[i_{\it nT}]=0$\\

\S bu6:/ \>$ pc\geq 100~\implies \quad \T busy/[i_{\it mig}]>0$
\end{tab1}

\medbreak Some other invariants we have postulated:

\begin{tab1} 
\S Ot1:/ \>$ \T X/(\B 0/)=\B null/$\\
\S Ot2:/ \>$ \T X/(a)\neq \B null/~\implies \quad \S ADR/(\T X/(a))=a$
\end{tab1}
The motivation of invariant (Ot1) is that we never store a value for 
the address 0.
The motivation of invariant (Ot2) is that the address in the hash table 
is unique. 

\begin{tab1} 
\S Ot3:/ \>$ return_{\it gA}=\{1, 10, 20, 30, 36, 46, 51\} \Land
return_{\it rA}=\{0, 59, 77, 90\}$ \\
\>$\land~ return_{\it ref}=\{10, 20, 30, 36, 46, 51\} \Land
return_{\it nT}=\{30, 46\}$ 
\end{tab1}

\begin{tab1} 
\S Ot4:/ \>$ pc\in \{$\=$0,~1,~5,~6,~7,~8,~10,~11,~13,~14,~15,~
16,~17,~18,~20,~$\\
\>\>$21,~25,~26,~27,~28,~30,~31,~32,~33,~35,~36,~37,~41,~$\\
\>\>$42,~43,~44,~46,~47,~48,~49,~50,~51,~52,~57,~59,~60,~$\\
\>\>$61,~62,~63,~65,~67,~68,~69,~70,~71,~72,~77,~78,~81,~$\\
\>\>$82,~83,~84,~90,~94,~95,~97,~98,~99,~100,~101,~102,~$\\
\>\>$103,~104,~105,~110,~111,~114,~116,~117,~118,~120,~$\\
\>\>$121,~123,~125,~126\}$
\end{tab1}

\section{Dependencies between invariants}
Let us write ``$\phi\ \B from/\ \psi_1, \cdots, \psi_n $'' to denote 
that $\phi$ is proved to be an invariant using that $\psi_1$, \dots, 
$\psi_n $ hold.
We write ``$\phi\ \Leftarrow\ \psi_1, \cdots, \psi_n $'' to denote 
that predicate $\phi$  is implied by the conjunction of  $\psi_1$, 
\dots, $\psi_n $.
We have verified the following ``\B from/'' and ``$\Leftarrow $'' 
relations mechanically:

\begin{tab}
Co1 \B from/ fi10, Ot3, fi1\\
Co2 \B from/ de5, Ot3, de6, del, de11\\
Co3 \B from/ in5, Ot3, in6, in1, in11\\
Cn1 \B from/ Cn6, Ot3\\
Cn2 \B from/ Cn8, Ot3, del\\
Cn3 \B from/ Cn10, Ot3, in1, in5\\
Cn4 \B from/ Cn11, Ot3\\
No1 $\Leftarrow $ No2\\
No2 \B from/ nT1, He2, rA2, Ot3, Ha2, Ha1, rA1, rA14, rA3, nT14, rA4\\
He1 \B from/ Ha1\\
He2 \B from/ Ha3, rA5, Ha1, He1, rA2\\
He3, He4 \B from/ Ot3, rA6, rA7, mi12, rA11, rA5\\
He5 \B from/ He1\\
He6 \B from/ rA8, Ha3, mi8, nT2, rA5\\
Ha1 \B from/ true\\
Ha2 \B from/ Ha1\\
Ha3 \B from/ Ha2, Ha1, He2, He1\\
Ha4 $\Leftarrow $ Ha3, He3, He4\\
Cn5 \B from/ Cn6, Ot3 \\
Cn6 \B from/ Cn5, Ot3\\
Cn7 \B from/ Cn8, Ot3, del\\
Cn8 \B from/  Cn7, Ot3 \\
Cn9 \B from/  Cn10, Ot3, in1, in5\\
Cn10 \B from/ Cn9, Ot3, in5\\
Cn11 \B from/ Cn11, Ot3\\
Cu1 \B from/ Ot3, Ha4, rA6, rA7, nT13, nT12, Ha2, He3, He4, rA11, nT9, nT10,\\
  \> mi13, rA5\\
Cu2 $\Leftarrow $ Cu6, cu7, Cu3, He3, He4\\
Cu3 \B from/ rA6, rA7, nT13, nT12, mi5, mi4, Ne8, rA5\\
Cu4 \= \B from/ del, in1, as1, rA6, rA7, Ha2, nT13, nT12, Ne9, Cu9, Cu10, de7, \\
  \> in7, as5, He3, He4, mi5, mi4, Ot3, Ha4, de3, mi9, mi10, de5, rA5\\
Cu6 \= \B from/ Ot3, rA6, rA7, Ha2, nT13, nT12, Ha3, in3, as3, Ne23, mi5, mE6,\\
  \> mE7, mE10, mE3, Ne3, mi1, mi4, rA5\\
Cu7 \= \B from/ Ot3, rA6, rA7, Ha2, nT13, nT12, Ha3, in3, as3, in5, mi5, mE6, \\
    \> mE7, mE10, mE3, Ne3, mi4, de7, in7, as5, Ne22, mi9, mi10, rA5, He3, \\
    \> mi12, mi1, Cu9, de1 in1, as1\\
Cu8 \= \B from/ Cu8, Ha2, nT9, nT10, rA6, rA7, mi5, mi4, mC2, mC5, He3,\\
    \> He4, Cu1, Ha4, mC6, mi16, rA5\\
Cu9\=, Cu10 \B from/ rA6, rA7, nT13, nT12, Ha2, He3, He4, Cu1, Ha4, de3, in3,\\
 \> as3, mE3, mi9, mi10, mE10, mE7, rA5\\
Cu11\=, Cu12 \B from/ Cu9, Cu10, Cu13, Cu14, del, in1, as1, rA6, rA7, Ha2, nT13,\\
 \> nT12, He3, He4, Cu1, Ha4, in3, as3, mi14, mi15, de3, in10, as8, mi12, \\
 \> Ot2, fi5, de8, in8, as6, Cu15, de11, in11, rA5\\
Cu13\=, Cu14 \B from/ He3, He4, Ot2, del, in1, as1, Ot1, rA6, rA7, nT13, nT12, \\
    \> Ha2, Cu9, Cu10, Cu1, Ha4, de3, in3, as3, Cu11, Cu12, in10, as8, fi5, de8, \\
    \> in8, as6, Cu15, mi17, mi18, mi12, mi4, de11, rA5\\
Cu15\=\ \B from/ He3, He4, rA6, rA7, nT13, nT12, Ha2, Cu1, Ha4, del, in1, as1,\\
    \> de3, in3, as3, fi5, de8, in8, as6, mi12, mi19, mi4, Ot2, Cu9, Cu10, Cu11, \\
    \> Cu12, Cu13, Cu14, rA5 \\
Cu16 $\Leftarrow $ Cu13, Cu14, Cu15, He3, He4, Ot1\\
Ne1 \B from/ nT9, nT10, mi7\\
Ne2 \B from/ Ne5, nT3, mi8, nT9, nT10\\
Ne3 \B from/ Ne1, nT9, nT10, mi8\\
Ne4 \B from/ Ne1, nT9, nT10\\
Ne5 \B from/ Ot3, nT9, nT10, mi5\\
Ne6 $\Leftarrow $ Ne10, Ne24, He6, He3, He4, Cu4\\
Ne7 \B from/ Ha3, rA6, rA7, rA8, nT13, nT12, nT11, He3, He4, mi8, nT7, \\
    \> Ne5, Ha2, He6, rA5\\
Ne8 \B from/ Ha3, rA8, nT11, mi8, nT6, Ne5, rA5\\
Ne9 \B from/ Ha3, Ha2, Ne3, Ne5, de3, as3, rA8, rA6, rA7, nT8, nT11, mC2, \\
    \> nT4, mi8, rA5\\
Ne9a \B from/ Ha3, Ne3, rA5, de3, rA8, nT4, mi8\\
Ne10 \B from/ Ha3, Ha2, de3, rA8, nT11, Ne3, He6, mi8, nT8, mC2, nT2, Ne5, \\
    \> rA5\\
Ne11 \B from/ Ha3, Ha2, He6, nT2, nT8, rA8, nT11, mi8, Ne3, mC2, rA5\\
Ne12\=, Ne13 \B from/ Ha3, Ha2, Cu8, He6, He3, He4, Cu1, de3, in3, as3, rA8, rA6, \\
    \> rA7, nT11, nT13, nT12, mi12, mi16, mi5, mi4, de7, in7, as5, Ot2, del,in1, \\
    \> as1, Cu9, Cu10, Cu13, Cu14, Cu15, as9, fi5, de8, in8, as6, mC2, Ne3, Ot1, \\
    \> Ne14, Ne20, mE16, mE7, mE4, mE1, mE12, mE2, Ne15, Ne16, Ne17,\\
    \> Ne18, mi20, de11, in11, rA5 \\
Ne14\=\ \B from/ Ha3, Ha2, He6, He3, He4, nT2, nT8, de3, in3, as3, rA8, nT11,\\
    \> Ot2, del, in1, as1, Cu9, Cu10, mi8, Ne3, mC2, mE7, mE16, mE1, mE4, \\
    \> mE12, Ne17, Ne18, Cu1, rA5\\
Ne15\=, Ne16 \B from/ Ha3, Ha2, Cu8, He6, He3, He4, Cu1, de3, in3, as3, rA8, rA6, \\
    \> rA7, nT11, nT13, nT12, mi12, mi16, mi5, mi4, de7, in7, as5, Ot2, del, in1, \\
    \> as1, Cu9, Cu10, Cu13, Cu14, Cu15, as9, fi5, de8, in8, as6, mC2, Ne3, Ot1, \\
    \> Ne19, Ne20, Ne12, Ne13, mE16, mE7, mE4, mE1, mE12, mE10, mE2, \\
    \> in11, de11, rA5\\
Ne17\=, Ne18 \B from/ Ha3, Ha2, mi8, He6, He3, He4, Cu1, nT2, de3, in3, as3, rA8,\\
    \> rA6, rA7, nT11, nT13, nT12, de7, in7, as5, Ot2, del, in1, as1, Cu9, Cu10, \\
    \> nT8, mE2, fi5, de8, in8, as6, mC2, Ne3, mC11, mC6, mC12, mE7, mE10,\\
    \> mE1, Cu8, Cu15, Cu13, Cu14, Cu11, Cu12, as8, de11, rA5\\
Ne19\=\ \B from/ Ha3, Ha2, He6, nT2, nT8, de3, in3, as3, rA8, nT11, mi8, Ne3, \\
    \> mE7, Ne14, mE16, Ot1, mE1, mE4, mE12, Ne17, Ne18, rA5\\
Ne20\=\ \B from/ Ha3, Ha2, Cu8, He6, He3, He4, Cu1, Ha4, de3, in3, as3, rA8, rA6,\\
    \> rA7, nT11, nT13, nT12, mi12, mi16, mi5, mi4, Ne1, de7, in7, as5, del, in1,\\
    \> as1, Cu9, Cu10, Cu13, Cu14, Cu15, as9, fi5, de8, in8, as6, mC2, Ne3, Ot1,\\
    \> mi20, in11, rA5\\
Ne22\=\ \B from/ Ot3, rA8, Ha2, nT11, Ha3, de3, in3, as3, mi5, mi4, Ne3, nT18,\\
    \> mE3, mi8, mE10, mE7, mE6, Ne5, nT5, nT2, rA5, nT8, nT12, mC2, mE2\\
Ne23 $\Leftarrow $ Cu6, cu7, Ne6, Ne7, He3, He4, Ne22, He6\\
Ne24 $\Leftarrow $ Ne27, He6 \\
Ne25 $\Leftarrow $ Ne19, Ne17, Ne18, He6 \\
Ne26 $\Leftarrow $ Ne17, Ne18, He6\\
Ne27 $\Leftarrow $ Cu16, Ne25, Ne26, Ne17, Ne18, He6 \\
fi1, del, in1, as1 \B from/ \\
fi2 \B from/ fi2, Ot3 \\
fi3 \B from/ fi4, Ot3, rA6, rA7, Ha2, rA5\\
fi4 \B from/ Ot3, rA6, rA7, nT13, nT12 \\
fi5, de8, in8, as6 $\Leftarrow $ Cu2, de10, in10, as8, fi8, He3, He4\\
fi6 \= \B from/ Ot3, fi1, del, in1, as1, rA6, rA7, Ha2, nT13, nT12, mi9, mi10, Cu9, \\
    \> Cu10, He3, He4, Cu1, Ha4, fi4, in3, as3, rA5\\
fi7 \= \B from/ fi8, fi6, fi2, Ot3, fi1, del, in1, as1, rA6, rA7, Ha2, nT13, nT12, mi9, \\
    \> mi10, Cu9, Cu10, He3, He4, Cu1, Ha4, fi4, in3, as3, rA5\\
fi8 \= \B from/ fi4, fi7, fi2, Ot3, fi1, del, in1, as1, rA6, rA7, Ha2, nT13, nT12, mi9,\\
    \> mi10, Cu9, Cu10, He3, He4, Cu1, Ha4, in3, as3, rA5\\
fi9 \= $\Leftarrow $ Cu1, Ha4, Cu9, Cu10, Cu11, Cu12, fi8, fi3, fi4, fi5, de8, in8, \\
    \> as6, He3, He4\\
fi10 \B from/ fi9, Ot3\\
fi11\=, de12, in12, as10 \B from/ Ot3, nT9, nT10, mi9, mi10, Cu8, fi4, de3, in3, \\
    \> as3, fi3, de2, in2, as2\\
de2 \B from/ de3, Ot3, rA6, rA7, Ha2, rA5\\
de3 \B from/ Ot3,  rA6, rA7, nT13, nT12 \\
de4, in4, as4 \B from/ Ot3\\
de5 \B from/ Ot3 \\
de6 \B from/ Ot3, de1, de11 \\
de7, in7, as5 $\Leftarrow $ de3, in3, as3, Cu1, Ha4, de13, in13, as11\\
de9 \= \B from/ Ot3, del, in1, as1, rA6, rA7, Ha2, nT13, nT12, mi9, mi10, \\
    \> Cu9, Cu10, de3, de7, in7, as5, rA5\\
de10\=\ \B from/ de3, de9, Ot3, del, in1, as1, rA6, rA7, Ha2, nT13, nT12, mi9, \\
    \> mi10, Cu9, Cu10, de7, in7, as5, He3, He4, rA5\\
de11 $\Leftarrow $ de10, de2, de3, He3, He4, Cu1, Ha4, Cu9, Cu10, Cu11, Cu12, fi5, \\
    \> de8, in8, as6 \\
de13, in13, as11 $\Leftarrow $ Ax2, de2, de3, de4, in2, in3, in4, as2, as3, as4\\
in2 \B from/ in3, Ot3, rA6, rA7, Ha2, rA5\\
in3 \B from/ Ot3, rA6, rA7, nT13, nT12\\
in5 \B from/ Ot3\\
in6 \B from/ Ot3, in1, in11\\
in9 \= \B from/ Ot3, del, in1, as1, rA6, rA7, Ha2, nT13, nT12, mi9, mi10, Cu9,\\
    \> Cu10, He3, He4, in3, de7, in7, as5, rA5\\
in10\=\ \B from/ in9, fi2, Ot3, del, in1, as1, rA6, rA7, Ha2, nT13, nT12, mi9, mi10, \\
    \> Cu9, Cu10, He3, He4, in3, de7, in7, as5, rA5\\
in11 $\Leftarrow $ in10, in2, in3, Cu1, Ha4, Cu9, Cu10, Cu11, Cu12, fi5, de8, in8, as6 \\
as2 \B from/ as3, He3, He4, Ot3, rA6, rA7, Ha2, rA5\\
as3 \B from/ Ot3, rA6, rA7, nT13, nT12 \\
as7 \= \B from/ Ot3, del, in1, as1, rA6, rA7, Ha2, nT13, nT12, mi9, mi10, Cu9, \\
    \> Cu10, as3, de7, in7, as5, rA5\\
as8 \= \B from/ as7, Ot3, del, in1, as1, rA6, rA7, Ha2, nT13, nT12, mi9, mi10, Cu9, \\
    \> Cu10, He3, He4, as3, de7, in7, as5, rA5\\
as9 $\Leftarrow $ as8, as2, as3, He3, He4, Cu1, Ha4, Cu9, Cu10, Cu11, Cu12, fi5, de8,\\
    \> in8, as6 \\
rA1 \B from/ Ha2\\
rA2 \B from/ Ot3\\
rA3 \B from/ Ot3, rA9, He2, He1, rA2, rA13 \\
rA4 \B from/ Ot3, nT14\\
rA5 \B from/ Ot3, rA1, rA2, Ha3, He2\\
rA6\=, rA7 \B from/ Ot3, nT13, nT12, nT14, rA11, mi4, bu2, bu3, Ha3, mi6, Ha2,\\
    \> He3, He4, He2, rA2\\
rA8 \B from/ Ot3, bu4, nT14, mi6, Ne2, mi5\\
rA9 \B from/ Ot3, Ha2, nT14, He1, He2\\ 
rA10 \B from/ Ot3\\
rA11 \B from/ Ot3, nT13, nT12, mi2\\
rA12 \B from/ Ot3, nT9, nT10\\
rA13 \B from/ Ot3, rA5\\
rA14 \B from/ Ot3, rA4, He1, rA2\\
nT1 \B from/ Ot3, pr5, Ha3, nT14, nT16, Ha2\\
nT2 \B from/ Ot3, nT14, Ha3, rA5\\
nT3 \B from/ Ot3, nT9, nT10\\
nT4 \B from/ Ot3, Ha3, de3, nT13, nT12, nT15, rA5\\
nT5 \B from/ Ot3, Ha3, in3, as3, nT13, nT12, nT15, nT18, mE3, mi4, rA5\\
nT6 \B from/ Ot3, nT13, nT12, nT14, Ha3, rA5\\
nT7 \B from/ Ot3, nT13, nT12, nT15, rA6, rA7, Ha2, mi9, mi10, nT14, Ha3, \\
    \> nT16, rA5\\
nT8 \= \B from/ Ot3, de3, in3, as3, nT13, nT12, nT15, nT18, mE3, mi4, Ha3, mC2,\\
    \> nT16, nT2, Ha2, rA5\\
nT9, nT10 \B from/ Ot3, pr2, pr3, nT18 \\
nT11 \B from/ Ot3, pr4, nT16, mi8\\
nT13, nT12 $\Leftarrow $ nT9, nT10, Ha3, He3, He4\\
nT14 \B from/ Ot3, nT9, nT10, nT18, nT16, pr7\\
nT15 $\Leftarrow $ nT14, Ha3, nT2\\
nT16 \B from/ Ot3, pr8\\
nT17 \B from/ Ot3, mi5, pr4, nT11, mi10\\
nT18 \B from/ Ot3, pr9, mi5, nT11\\
mi1 \B from/ Ot3, mi9, mi10, mi10\\
mi2 \B from/ Ot3, Ne4\\
mi3 \B from/ Ot3, fi11, de12, in12, as10, nT9, nT10, Ne5\\
mi4 \B from/ Ot3, mi9, mi10, mi3\\
mi5 \B from/ Ot3, nT9, nT10, Ne5, mi10, mi4\\
mi6 \B from/ Ot3, mi5, bu6, rA8, mi9, mi10, bu4, mi4\\
mi7 \B from/ Ot3, mi2, mi7, mi4, nT18, Ne2, mi10, nT17, mi3\\
mi8 \B from/ Ot3, mi10, Ne2, mi3\\
mi9,\=\ mi10 \B from/ Ot3, He3, He4, nT9, nT10, nT18, Ne3, Ha3, mi3, nT17, \\
    \> mi10, He2, mi4, mi12, mi6, He6 \\
mi11 \B from/ Ot3, nT18, mi9, mi6, mi6\\
mi12 \B from/ Ot3, rA8, nT2, He6, mi9, mi5, mi3, Ha3, mi4, rA5\\
mi12 \B from/ Ot3, mi12, nT18, mi6, Ha3, mi4, rA5\\
mi13 \B from/ Ot3, rA6, rA7, Ha2, nT13, nT12, He3, He4, mi9, mi10, mC9, rA5\\
mi14\=, mi15 $\Leftarrow $ Ne12, Ne13, mi5, Cu15, mi13, Ot2, He3, He4, Ne17, Ne18,\\
    \> Cu8, He6, He5, mi4, Ot1\\ 
mi16 $\Leftarrow $ Ne11, mi5, mi4 \\
mi17, mi18 $\Leftarrow $ Ne15, Ne16, mi5, Cu15, mi13, Ot2, He3, He4, Ne17, Ne18, \\
    \> Cu8, He6, He5, mi4\\
mi19 $\Leftarrow $ Ne20, mi5, Cu15, mi13, Ot2, He3, He4\\
mi20 \=\ \B from/ Ha3, Ha2, Cu8, He6, He3, He4, Cu1, Ha4, de3, in3, as3, rA8, \\
     \> rA6, rA7, nT11, nT13, nT12, mi5, mi4, de7, in7, as5, Ot2, del, in1, \\
     \> as1, Cu9, Cu10, Cu13, Cu14, Cu15, as9, fi5, de8, in8, as6, mC6, Ne3,\\
     \> Ot3, mC11, mi13, mi9, mi10, mC2, mE3, mE10, mE7, mC12, mE1, \\
     \> mE13, Ne17, Ne18, mE2, mE4, Ot1, mE6, Ne10, in11, rA5\\
mC1 \B from/ Ot3, mi6, mi11, nT18 \\
mC2 \B from/ Ot3, rA6, rA7, nT13, nT12, mC2\\
mC3 \B from/ Ot3, mC3, nT13, nT12, rA6, rA7, Ha2, rA5\\
mC4 \B from/ Ot3, mC4, mC2, mC3, He3, He4, rA6, rA7, Ha2, rA5\\
mC5 \B from/ Ot3 \\
mC6 \B from/ Ot3, rA6, rA7, Ha2, nT13, nT12, mC2, rA5\\
mC7 \B from/ Ot3, rA6, rA7, Ha2, nT13, nT12, mC2, rA5\\
mC8 \B from/ Ot3, rA6, rA7, Ha2, nT13, nT12, He3, He4, mC7, rA5\\
mC9 \B from/ Ot3, rA6, rA7, Ha2, nT13, nT12, He3, He4, mi9, mi10, He5, \\
    \> mC7, mC8, rA5\\
mC10\=\ \B from/ Ot3, rA6, rA7, Ha2, nT13, nT12, mC2, del, in1, as1, mi6,\\
    \> Ha3, mi4, nT18, mE15, mC11, mi5, rA5\\
mC11 \B from/ Ot3, rA6, rA7, Ha2, nT13, nT12, mC2, rA5\\
mC12 \B from/ Ot3, rA6, rA7, mC2, mC11, Cu9, Cu10, de7, in7, as5, mi9,\\
     \> mC6 \\ 
mE1 \B from/ Ot3 \\
mE2 \B from/ Ot3 \\
mE3 \B from/ mC1, Ot3, mi6, nT18 \\
mE4 \B from/ Ot3, mE1\\
mE5 \B from/ Ot3, mE3, Ha3, mi6, mi4, nT18, Ha2, rA5\\
mE6 \B from/ Ot3 \\
mE7 \B from/ Ot3, Ha2, Ha3, mi6, mi4, mE3, rA5\\
mE8 \B from/ Ot3, Ha3, mi6, mi4, nT18, Ha2, mE3, rA5\\
mE9 \= \B from/ Cu1, Ha4, Ot3, Ha2, Ha3, mi6, mi4, mE3, mC2, mC10, mE1, \\
    \> mC1, del, in1, as1, mi13, mi12, mC6, mE2, rA5\\
mE10 \B from/ del, in1, as1, mE3, mi6, Ot3, Ha2, Ha3, mi4, mE11, mE9,\\
    \> mE7, rA5\\
mE11\=\ $\Leftarrow $ mE10, mi13, mE16, mi16, mi5, mE3, Ne12, Ne13, mC12, mE2,\\
    \> mE1, mE4, mC6, mE12, mi12, Cu13, Cu14, He3, He4, mi4\\
mE12 $\Leftarrow $ Ne23, Ne22, mE16, He6, Ne8\\
mE13 \B from/ Ot3, Ha2, mE14, del, in1, as1, Ha3, mi6, mi4, mE3, rA5\\
mE14 \B from/ Ot3, Ha2, del, in1, as1, Ha3, mi6, mi4, nT18, mE3, mE2, rA5\\
mE15\=\ \B from/ Ot3, mE1, Ha2, del, in1, as1, Ha3, mi6, mi4, nT18, mE3, mE2,\\
    \> mE7, mE14, mE4, rA5\\
mE16\=\ \B from/ Ha3, Ha2, mE3, del, in1, as1, mi6, mE2, mE4, mE1, mE7, mi4,\\
    \> Ot3, mE14, mE13, rA5\\
pr1 \B from/ Ot3, rA11, rA10, nT9, nT10, Ne5, mi2, mi4, mi8, mi5\\
pr2, pr3 \B from/ pr1, Ot3, rA11, mi1 \\
pr4 $\Leftarrow $ pr1\\
pr5 $\Leftarrow $ pr6, pr1, bu1\\
pr6 \B from/ Ot3, Ha2, nT9, nT10, nT14, nT16, He2, rA2, pr1, bu1, pr10, \\
    \> rA9, He1, rA4\\
pr7, pr8, pr9 $\Leftarrow $ pr1, mi4\\
pr10 \B from/ Ot3, pr1, nT9, nT10, nT14, nT17\\
bu1 \B from/ Ot3, rA11, rA10, nT9, nT10, Ne5, mi2, mi8, mi5, bu5\\
bu2, bu3 $\Leftarrow $ bu1, Ot3, rA10\\
bu4 $\Leftarrow $ bu1\\
bu5 \B from/ Ot3, nT9, nT10, nT16, nT18, pr1, bu1\\
bu6 $\Leftarrow $ bu1, mi4\\
Ot1 \B from/ del, in1, as1\\
Ot2 \B from/ del, in1, as1\\
Ot3 \B from/ true\\
Ot4 \B from/ Ot3\\
\end{tab}

\end{document}